\documentclass[aip,pof,superscriptaddress,reprint,onecolumn]{revtex4-1}

\draft

\usepackage{graphicx}
\usepackage{epstopdf}
\usepackage{epsfig,bm,epsf,float}
\usepackage{lineno}
\usepackage[colorlinks=true,linkcolor=black,anchorcolor=black,citecolor=black,filecolor=black,menucolor=black,runcolor=black,urlcolor=black]{hyperref}
\usepackage[utf8]{inputenc}
\usepackage{dcolumn}
\usepackage{bm}
\usepackage{amsfonts}
\usepackage{amsmath}
\usepackage{amssymb}
\usepackage{amsbsy}
\usepackage{psfrag}
\usepackage{color}
\usepackage{soul}
\usepackage[normalem]{ulem}
\usepackage{stackengine}
\usepackage{float}
\usepackage{makecell}
\usepackage{comment}
\usepackage{multirow}
\usepackage[dvipsnames]{xcolor}
\usepackage{placeins}

\graphicspath{{figures/0-intro},{figures/1-rectilinear/},{figures/2-zigzagging/},{figures/3-spiral/},{figures/4-comparison/},{figures/5-short_wall/}}

\newlength{\subcolumnwidth}

\newcommand{\nextsubcolumn}[1][]{
  \cr\noalign{\hfill}
  \if\relax\detokenize{#1}\relax\else\hsize=#1\setlength{\subcolumnwidth}{\hsize}\fi
}

\begin{document}

\preprint{Manuscript in preparation}

\title{Bubble rising near a vertical wall: experimental characterization of paths and velocity.}

\author{C. Estepa-Cantero}
\email{cestepa@ugr.es}
\affiliation{\'Area de Mec\'anica de Fluidos, Departamento de Mec\'anica de Estructuras e Ingenier\'{\i}a Hidr\'aulica, Universidad de Granada, Campus Fuentenueva s/n, 18071, Granada, Spain.}
\affiliation{Andalusian Institute for Earth System Research, University of Granada. Avda. del Mediterr\'aneo s/n, 18006, Granada, Spain.}
\author{C. Mart\'inez-Baz\'an}
\email{cmbazan@ugr.es}
\affiliation{\'Area de Mec\'anica de Fluidos, Departamento de Mec\'anica de Estructuras e Ingenier\'{\i}a Hidr\'aulica, Universidad de Granada, Campus Fuentenueva s/n, 18071, Granada, Spain.}
\affiliation{Andalusian Institute for Earth System Research, University of Granada. Avda. del Mediterr\'aneo s/n, 18006, Granada, Spain.}
\author{R. Bola\~nos-Jim\'enez}
\email{rbolanos@ujaen.es}
\affiliation{\'Area de Mec\'anica de Fluidos, Departamento de Ingenier\'{\i}a Mec\'anica y Minera. Universidad de Ja\'en. Campus de las Lagunillas, 23071, Ja\'en, Spain.}
\affiliation{Andalusian Institute for Earth System Research, Universidad de Ja\'en. Campus de las Lagunillas, 23071, Ja\'en, Spain.}

\date{\today}

\begin{abstract}
The trajectories of a single bubble rising in the vicinity of a vertical solid wall are experimentally investigated. Distinct initial wall-bubble distances are considered for three different bubble rising regimes, i.e. rectilinear, planar zigzag, and spiral. The problem is defined by three control parameters, namely the Galilei number, $Ga$, the Bond number, $Bo$, and the initial dimensionless distance between the bubble centroid and the wall, $L$. We focus on high-Bond numbers, varying $L$ from 1 to 4, and compare the results with the corresponding unbounded case, $L \rightarrow \infty$. In all cases, the bubble deviates from the expected unbounded trajectory and migrates away from the wall as it rises due to the overpressure generated in the gap between the bubble and the wall. This repulsion is more evident as the initial wall-bubble distance decreases. Moreover, in the planar zigzagging regime, the wall is found to impose a preferential zigzagging plane perpendicular to it when $L$ is small enough. Only slight wall effects are observed in the velocity or the oscillation amplitude and frequency. The wall migration effect is more evident for the planar zigzagging case and less relevant for the rectilinear one. Finally, the influence of the vertical position of the wall is also investigated. When the wall is not present upon release, the bubbles have the expected behavior for the unbounded case and experience the migration only instants before reaching the wall edge. This repulsion is, in general, more substantial than in the initially-present-wall case.
\end{abstract}

\maketitle

\section{Introduction}\label{sec:intro}
Bubble rising in liquids constitutes a relevant two-phase flow phenomenon that has been the subject of vast research for decades. Its paramount importance relies not only on its wide application in many different industrial operations but also on its broad occurrence in numerous natural processes, such as methane seep in marine fissures or gas bubbles in volcano eruptions~\citep{dean2018methane,liu2015role}. Regarding technical applications, bubble rising is relevant in areas as diverse as water treatment, mineral flotation, carbon capture, drag reduction, ocean microplastics scavenging, or medical treatment and diagnosis, among many other techniques~\citep{Rodriguez-Rodriguez2015,rajapakse2022effects,Shaw2022,Wang2022,Zhao2023,Lehmann2023,Gao2023,Ri2023}.

The rising of an isolated bubble in an unbounded configuration has been thoroughly investigated in the past~\citep{Clift1978,marusic2021leonardo,Yang2003} not only theoretically and numerically~\citep{blanco1995structure,Magnaudet2000,mougin2002paht,mougin2006wake,magnaudet2007wake,tripathi2015dynamics,Cano-Lozano2015,Cano-Lozano2016,Zhang2021,bonnefis2023and}, but also experimentally~\citep{haberman1953experimental,Bhaga1981,Duineveld1995,Maxworthy1996,Magnaudet2000,Ellingsen2001,DeVries2003,Shew2006,zenit2008path,veldhuis2008shape,Legendre2012,Aoyama2016,Rastello2017,agrawal2021experimental,kure2021experimental,She2021,liu2022experimental}. However, most of these studies are focused on bubbles at low-Bond and high-Reynolds numbers, with the Bond number $Bo=\rho g D^{*2}/\sigma$ and the Reynolds number $Re=\rho v^* D^*/\mu$, where $\rho$ and $\mu$ the liquid density and viscosity, respectively, $\sigma$ the surface tension, $g$ gravity acceleration, $D^*$ the equivalent bubble diameter and $v^*$ the bubble terminal velocity. A general outcome of these works is that the path that the bubble follows as it rises in a quiescent liquid depends on Bond, $Bo$, and Galilei, $Ga$, numbers, with $Ga=\rho g^{1/2} D^{* 3/2}/\mu$, and therefore on the bubble shape (or the major-to-minor bubble diameter ratio, $\chi$). At low $Bo$, the bubble is spherical and, apart from a standing eddy on the rear, no vortical structures are developed. Thus, it tends to follow a rectilinear path~\citep{Clift1978,blanco1995structure,Cano-Lozano2016}. As $Bo$ increases, the bubble aspect ratio also increases, and the bubble wake may become unstable, leading to path instability~\citep{Mougin2002,magnaudet2007wake}, and providing two-dimensional (planar zigzagging regime) or even three-dimensional (spiraling or flat-spiraling regimes) paths~\citep{Yang2003,zenit2008path}. The planar zigzagging path is characterized by the shedding of two or four symmetric counter-rotating trailing vortices\citep{Cano-Lozano2016,She2021}, while the wake of the spiraling bubble shows two intertwined vortices wrapping around each other\citep{Cano-Lozano2016}.

However, there are many real situations where the bubbles do not freely rise but they unavoidably interact with solid boundaries. Under these conditions, certainly, their rising dynamics are modified by the presence of the wall. Many theoretical and numerical works have addressed this problem~\citep{Magnaudet2003,magnaudet2007wake,sugiyama2010lateral,sugioka2015direct,Zhang2020,Hasan2022} reporting a clear migration away from the wall, caused by the constrain imposed in the bubble movement.  Indeed, \citet{Magnaudet2003} performed an analytical study to describe the deformation-induced lift force on the bubble. As for numerical works, \citet{sugiyama2010lateral} corroborated that the migration of the bubble away from the wall happened due to the deformation of the bubble. They showed that the theoretical approaches available at that time did not provide an accurate prediction of the amplitude of this migration for the smallest bubble-wall distances and presented a new approach to predict the motion of slightly deformed bubbles under these conditions. \citet{Zhang2020} performed unsteady, three-dimensional numerical simulations to describe the dynamics of rectilinear and spiraling rising bubbles in the vicinity of a vertical wall. They stated that the wall induced a destabilizing effect in the wall-normal direction while it acted as a stabilizing factor in the spanwise direction. They also observed an increase in the bubble-wall separation and in the lateral oscillations which was stronger the closer the bubble was initially to the wall. They found that the wall caused a lateral disturbance whose magnitude increased as the wall-bubble initial distance decreased, although these results have not been experimentally confirmed yet. Regarding the spiraling motion, \citet{Zhang2020} observed that the phase was modified by the wall, but the amplitude and frequency of oscillation were both unaffected by the initial distance. Moreover, \citet{Yan2022} reported numerically that a greater initial bubble-wall distance promoted less deflection and that the path oscillations leading to an eventual spiral motion began earlier as the distance to the wall decreased. In addition, they observed that the bubble terminal velocity was slightly smaller than its unbounded counterpart due to the wall additional viscous effect, which leads to an increase in the bubble drag force.
\citet{Hasan2022} studied the migration dynamics of a deformable bubble in the presence of not one wall, but two of them forming a corner. They reported that the bubble suffered a migration perpendicular to the corner, i.e. forming 45$^o$ with each surface, because its motion was constrained in two directions. They also concluded that the wall was a destabilizing factor on the wall-normal direction and attributed the strongest deformations of the bubble to the modification of wake structure generated by the walls.
Finally, \citet{yan2023three} related the lateral migration to changes in the flow distribution around the bubble by means of simulations. They claimed that the wall presence favors bubble oscillations along the wall-normal plane, being restricted across the parallel direction. Moreover, they concluded that the wall promotes bubble vortex shedding, and therefore the transition from a stable rising regime to an unstable one.

The migration of bubbles rising in the presence of a vertical wall has been hardly explored experimentally due to the complexity involved in performing these kinds of experiments in a controlled manner. Still, a few experimental works \citep{Takemura2002,Takemura2003,Jeong2015,Lee2017,Yan2022,yan2023three} can be found, although at  low-Bond regimes. In particular, \citet{Takemura2002} and \citet{Takemura2003} investigated the motion of a spherical bubble near a wall for $Re<100$ and reported that the repulsion force acting on the bubble happens due to the asymmetric distributions of the vortices with respect to the bubble symmetry plane. 
Their work was corroborated numerically by \citet{sugioka2015direct} with only slight differences and they clarified that the drag force increased with the bubble-wall proximity. Moreover, \citet{Jeong2015} tested a zigzagging bubble, finding that the bubble mostly bounced off the wall for distances smaller than the bubble diameter. Under very similar conditions, \citet{Lee2017} used Particle Image Velocimetry to experimentally investigate the wake behind a zigzagging bubble in the proximity of a wall.

The objective of the present work is to improve the knowledge on the rising of bubbles near a wall by extending our study to regimes of high Bond numbers, which have barely been explored experimentally. These millimeter-sized bubbles are present in many natural and industrial processes, such as in aeration tanks used in bioreactors and chemical reactors, or mineral purification by flotation techniques, among many others. Furthermore, the experimental results reported here might be useful to corroborate the numerical and theoretical approaches.

This work is organized as follows. Section \ref{sec:exp} describes the experimental facility and the methodology followed. Section \ref{sec:results} reports the results obtained for the three distinct regimes tested for different values of initial wall-bubble distance. A comparison of the wall effect among different regimes is included as well as the impact of the vertical position of the wall on bubble trajectory. Finally, section \ref{sec:conclusions} is devoted to the conclusions extracted from our analysis.
\begin{figure}[t!]
    \centering
    \includegraphics[width=.8\textwidth]{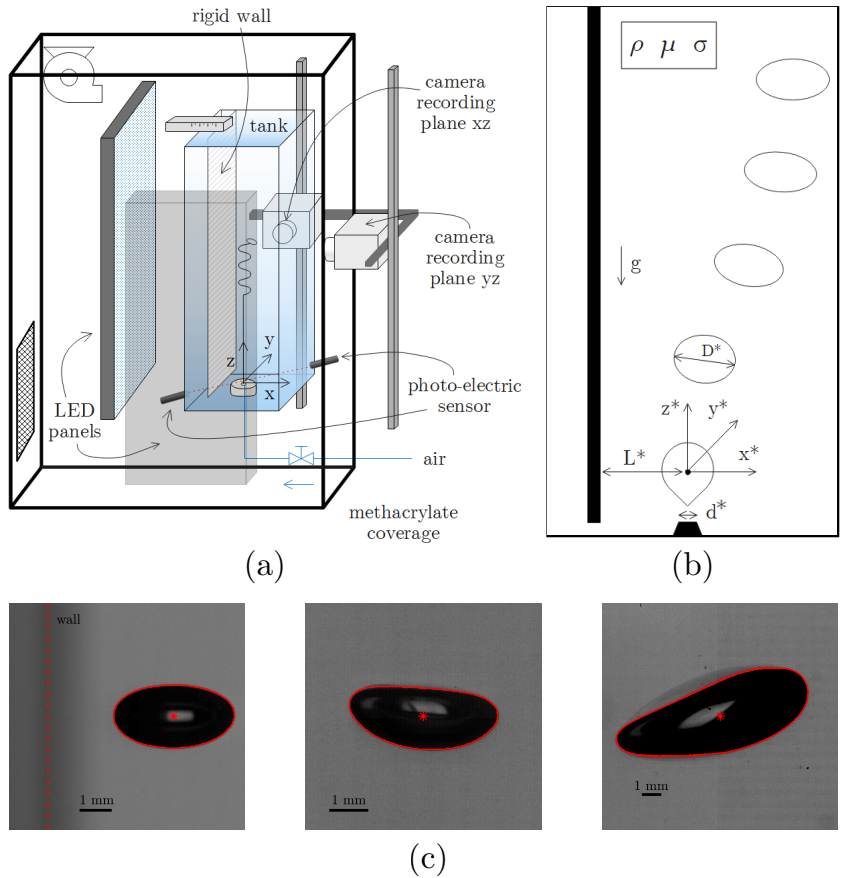}
    \caption{(a) Sketch of the experimental setup. (b) Sketch of the physical problem including the physical and geometrical governing parameters. Note that the reference-frame origin is set at the bubble centroid right after bubble pinch-off. The trajectory is vertically defined by $z^*$ while $x^*$ and $y^*$ are the horizontal axes in the direction perpendicular and parallel to the wall, respectively. (c) Bubble images corresponding to the three regimes studied here, taken from the high-speed recordings, showing the bubble contour and their centroids (red line and asterisk, respectively). From left to right: rectilinear, planar zigzag and spiral regimes.}
    \label{fig:setup}
\end{figure}

\section{Experimental setup and methods}\label{sec:exp}
\begin{figure}[t!]
    \centering
    \includegraphics[width=\textwidth]{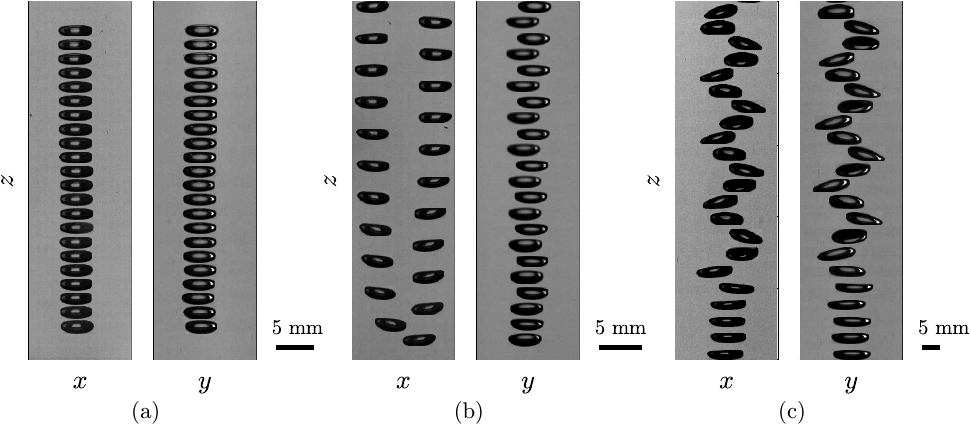}
    \caption{Images of an individual bubble at different instants, in the $xz$ (left panel) and $yz$ (right panel) planes, extracted from the high-speed recordings for the unbounded case and the a) rectilinear, b) zigzagging and c) spiral regimes. Note that small oscillations can be observed in the plane $yz$ in (b) because the views shown do not correspond with the zigzagging plane, see section \ref{subsec:zig}.}
    \label{fig:prueba}
\end{figure}
The experimental setup is shown in Fig.~\ref{fig:setup}(a). It consisted of an open tank 1.2 m high with a square cross-section 0.13 m wide. To assure optical access and cleanliness, the four sides of the tank were constructed of glass and a methacrylate cabin covered the facility entirely. Bubbles were generated injecting a given flow rate of air through an injector placed at the center of the tank base. Injectors of different diameters, ranging from 2 to 12 mm, were used to generate bubbles of different sizes (see Table~\ref{tab:cases}). A capillary tube was inserted between the injector and the air feeding line to guarantee constant flow rate conditions, and very small air flow rates were supplied to ensure quasi-static bubble formation. Indeed, high-speed video images of the bubble formation were analyzed to check this condition. Finally, a glass wall was placed vertically inside the tank, parallel to the $yz$ plane, to study its effect on the motion of the rising bubble. The horizontal distance from the wall to the center of the injector was controlled with a rack and pinion mechanism with micrometric precision. 
 \begin{table}[t!]
 \caption{Main characteristics of the three cases studied in this work. $^\dagger$$d^*$=12 mm is the diameter of a circular superhydrophobic surface used as an air injector in this case. $^\ddagger$ $St=$ 0.132$\pm$0.006 for wall-bounded cases and $St=$0.160$\pm$0.007 for the free case. All tabulated data are stated as average values $\pm$ standard deviation of all the experiments from both bounded and unbounded cases and the three regimes, except $St$ for the spiral regime, that exhibits different values for bounded and free case.}\label{tab:cases}
\centering
\resizebox{\textwidth}{!} {
\begin{tabular}{c c c c c c c c c}
    \hline\hline
    Case & Regime & Liquid & $d^*$ (mm) & $D^*$ (mm) & $Bo$ & $Ga$ & $Mo$ ($\times 10^{6}$) & $St$\\ [0.5ex]
    \hline
    1 & Rectilinear & T11 & 2.50 & 3.31\textpm0.07 & 5.0\textpm0.2 & 59.7\textpm1.8 & 9.87 & -\\ 
    \hline
    2 & Planar zigzag & T05 & 2.00 & 2.92\textpm0.01 & 3.87\textpm0.03 & 98.7\textpm0.6 & 0.61 & 0.109\textpm0.004\\
    \hline
    3 & Spiral & GW & 12.00$^\dagger$ & 7.62\textpm0.07 & 10.29\textpm0.15 & 108\textpm8 & 8.4\textpm2.3 & \begin{tabular}{@{}c@{}}0.132\textpm0.006$^\ddagger$ \\ 0.160\textpm0.007$^\ddagger$\end{tabular}\\
    \hline\hline
 \end{tabular}
 }
 \end{table}

Figure~\ref{fig:setup}(b) shows a schematic representation of the problem addressed in this work. The reference-frame origin was set at the bubble centroid just after the bubble pinch-off. Directions $x^*$ and $y^*$ define the horizontal components of the bubble trajectory in the direction perpendicular and parallel to the wall respectively, while $z^*$ defines the vertical component. The variables involved in the free rise of a bubble are the equivalent bubble diameter, $D^*$, the liquid physical properties, $\rho,~\mu,~\sigma$, i.e. density, dynamic viscosity, and surface tension, respectively, and the gravity acceleration, $g$. The air density and viscosity are much smaller than those of the liquid, and thus they have been neglected. Since the bubbles in this study are deformable, the bubble size was characterized using the equivalent diameter $D^*$, defined as the diameter of a sphere with a volume, $V^*$, identical to that of the bubble, $D^*=(6V^*/\pi)^{1/3}$. Here asterisks denote dimensional parameters, in contrast to their non-dimensional counterparts. Taking $D^*$, gravitational velocity, $\sqrt{gD^*}$, and gravitational time, $\sqrt{D^*/g}$, as the characteristic length, velocity, and time, respectively, to obtain the dimensionless parameters for our analysis, the unbounded problem is fully defined by Bond, $Bo=\rho g D^{*2}/\sigma$, and Galilei, $Ga=\rho \sqrt{gD^{*3}}/\mu$, numbers~\citep{Cano-Lozano2016}. Morton number, $Mo=g\mu^4/\rho\sigma^3$, related to Bond and Galilei numbers as $Mo=Bo^3/Ga^4$, is also useful to describe this problem since it only depends on the liquid properties. The values of $Bo$, $Ga$, $Mo$, as well as the bubble diameters, corresponding to the experiments reported in this work, are summarized in Table \ref{tab:cases}. In addition, in the bounded case, the initial horizontal distance between the bubble centroid and the rigid wall, $L=L^*/D^*$, represents a relevant parameter. Our experiments were performed for four values of $L$, i.e. $L=1$, 2, 4 (wall-bounded cases) and $L \rightarrow \infty$ (unbounded case). Reynolds, $Re=\rho v^* D^*/\mu=Ga\,v$, and Weber, $We=\rho v^{*2} D^*/\sigma=Bo\,v^2$, numbers will be outcomes of the problem for each pair of ($Ga,Bo$) values, being $v^*$ the bubble terminal velocity and $v=v^*/\sqrt{gD^*}$ its dimensionless counterpart. Note that the latter is indeed the Froude number, $v=Fr$. Additionally, the unstable regimes exhibit oscillations with certain amplitude $A^*$ and frequency $f$, characterized by the dimensionless amplitude, $A=A^*/D^*$, and the Strouhal number, $St=f D^*/v^*$, respectively.
\begin{table}[!t]
\centering
\caption{Main properties of the three liquids used in this work. The values corresponding to the silicon oils have been provided by the manufacturer.}\label{tab:liquids}
\begin{tabular}{c c c c c c c c c}
    \hline\hline
    Liquid & Characteristics & $\rho$ (kg/m$^3$) & $\mu$ (mPa$\cdot$s) & $\sigma$ (mN/m)\\ [0.5ex]
    \hline
    T11 & Silicon oil & 935 & 9.35 & 20.1\\ 
    \hline
    T05 & Silicon oil & 913 & 4.57 & 19.7\\
    \hline
    GW & Glycerol-water & 1188.7\textpm1.1 & 22.1\textpm1.9 & 65.7\textpm0.1\\
    \hline\hline
 \end{tabular} 
 \end{table}
 
Three cases are considered in this work, each one exhibiting different rising paths according to the numerical study by \citet{Cano-Lozano2016}. In particular, we selected the liquids and air injector diameters in order to achieve conditions similar to those of bubbles 22, 19, and 26 in Table I of \citet{Cano-Lozano2016}. The three selected bubbles exhibit rectilinear, planar zigzagging, and spiraling paths respectively, all of them being in the proximity of the transition curves between regimes~\citep{Cano-Lozano2013,Cano-Lozano2016a}. The properties of the liquids used in this work are shown in Table \ref{tab:liquids}. Two different viscous silicon oils, T11 and T05 (Dow Corning$^{\circledR}$ XIAMETER\texttrademark\,PMX-200) were used for the liquid phase, similarly to the reference work. However, in Case 3, the capillary length associated with T11, the silicon oil considered for this case in the numerical study, was much smaller than the air injector radius required to generate the desired bubble size, preventing bubbles from being generated. In order to overcome this issue, a glycerol-water (GW) mixture was employed together with a superhydrophobic substrate (Rustoleum$^{\circledR}$ NeverWet\texttrademark) coating the air injector. The necessary diameter of the superhydrophobic substrate was obtained using the correlation given by \citet{Rubio-Rubio2021}. The GW mixture composition was 74.16-74.89\% in weight of glycerol, depending on the case. Distilled water was used to make the GW mixtures. The effects of surfactants on the bubble dynamics have been shown to be negligible when the bubble is large enough~\citep{Duineveld1995,Ellingsen2001,Shew2006}, like those considered in Case 3, as will be shown later. The physical properties of the mixtures were measured in the laboratory and compared to theoretical data \citep{dorsey_glyc-water,glycerine,Streit_glyc,Spann_glyc}. The temperature was continuously registered and the fluid properties were tested for the whole range of temperature during the experiments using a Brookfield DV3TLVCJ0 rheometer, a Krüss K20 force tensiometer, and a Mettler Toledo Density2Go Densito densimeter. The average values of the GW physical properties are shown in Table \ref{tab:liquids}. The corresponding values of the resulting dimensionless control parameters for the three experimental cases are summarized in Table~\ref{tab:cases}.

Before running any set of experiments, several steps were taken to ensure optimal experimental conditions. The tank and the vertical wall were thoroughly cleaned between different experimental runs using water and ethanol to eliminate dust and surfactants, which may change the dynamics of the problem~\citep{Duineveld1995}. The liquids were replaced after some experimental runs. To ensure the verticality of both the wall and the camera motion system, a plumb bob was hung inside the tank using a nylon monofilament cord with a calibrated diameter of 0.45 mm. The cord was recorded along the height of the tank to eliminate possible minor horizontal displacements of the system in the image processing. Moreover, before running each experiment, the bubble terminal velocity was determined. Afterwards, a servo motor was used to move the cameras similarly to previous works\citep{kure2021experimental}, being controlled with a SMC Lecsa2-S4 driver and the software Melsoft MR Configurator, which enabled us to program the cameras position, velocity, and acceleration over time.

The ascent of the bubble was filmed using two synchronized high-speed cameras (Fastcam SA1.1, Fastcam Mini Ax200) positioned perpendicularly to each other, and mounted on a traverse rail that moved at a pace similar to the bubble rising velocity. Thus, each camera recorded images from one side of the tank, parallel and perpendicular to the glass wall, to allow for a three-dimensional reconstruction of the bubble path. The cameras were also synchronized with the servomotor that controlled the motion of the traverse they were attached to in order to determine the bubble vertical position from the images. A laser beam and a photodiode sensor were used to detect the bubble pinch-off and to trigger the high-speed cameras recording and the traverse motion. A couple of 1.2-meter-long LED panels were used for backlighting, providing a highly uniform background. The shutter speed was adjusted depending on the camera model and the lighting conditions between 1/1000 and 1/30000 s. Every individual experimental case was performed at least 10 times to ensure repetitiveness. The time between each recorded bubble was long enough to avoid possible effects of the previous bubble wake on the results.

Examples of the evolution of the bubbles along their paths are shown in Fig.~\ref{fig:prueba} for the rectilinear, zigzagging and spiral regimes. The bubble contour and centroid were detected in each image using an in-house image processing routine based on the Matlab Image Segmentation Toolbox and an integral image thresholding method \citep{Otsu1979,Bradley2007}, as shown in Fig.~\ref{fig:setup}(c). The contours were used to compute the bubble size, shape, and equivalent diameter $D^*$, while the centroid position was used to obtain bubbles trajectories and velocities. The servomotor software allowed us to determine the $z$ coordinate of the centroid with an accuracy of 1 $\mu m$. The overall accuracy of the experiments depended on the temporal and spatial resolution of the films, which ranged from 500 to 2000 frames per second, and between 17.79 and 36.31 $\mu$m per pixel respectively. The error of the centroid position determined by image processing was evaluated using the estimation by \citet{Ho1983}. The absolute errors corresponding to different calculated parameters along the manuscript were estimated by applying propagation of uncertainty. 

\begin{figure}[ht!]
    \centering
    \includegraphics[width=.7\textwidth]{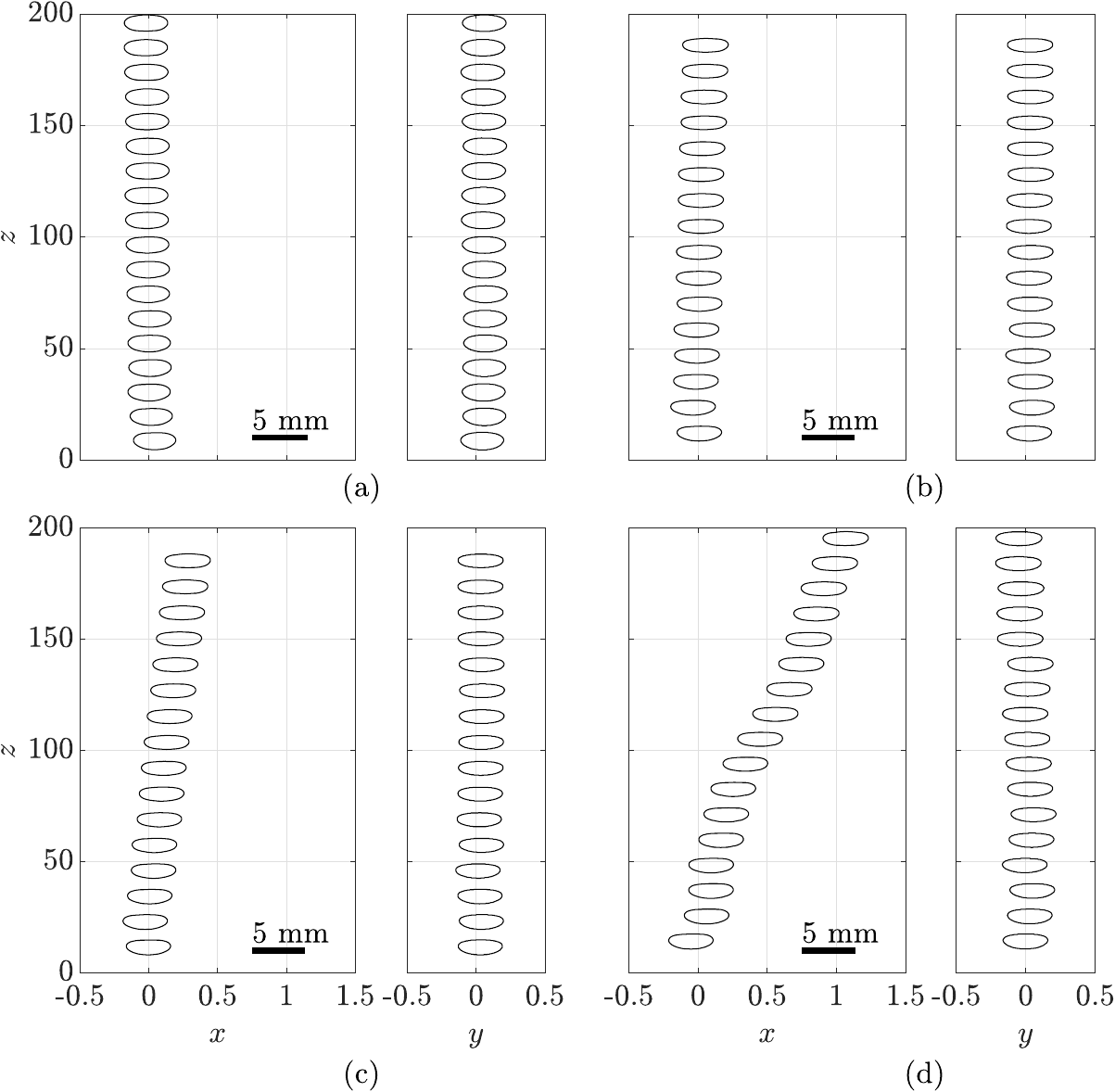}\vspace{-2ex}
    \caption{Bubbles silhouettes, extracted from the high-speed recordings at different instants of time as they rise, for the rectilinear regime in the $xz$ (left) and $yz$ (right) planes. (a) $L\rightarrow \infty$, (b) $L=4$, (c) $L=2$, and (d) $L=1$. The wall, not shown for the sake of clarity, is placed on the left-hand side of the $xz$ plane at (b) $x=-4$, (c) $x=-2$, and (d) $x=-1$. The bubbles have been scaled down for clarity.}
    \label{fig:contours_rect}
\end{figure}

\section{Results}\label{sec:results}
 This section is devoted to presenting the results obtained from the experiments. The section is organized as follows: the particular results corresponding to the rectilinear, planar zigzag, and spiral regimes are described in subsections~\ref{subsec:rect}, ~\ref{subsec:zig} and ~\ref{subsec:spi}, respectively. Subsection~\ref{subsec:comp} presents a direct comparison between the three regimes. Finally, the effect of the vertical distance from the wall to the injector is explored in subsection~\ref{subsec:wall}.

\subsection{Rectilinear regime} \label{subsec:rect}
Stable bubbles that rise with a rectilinear path were obtained using T11 silicon oil together with a 2.5 mm diameter injector (Case 1 in Table~\ref{tab:cases}). The resulting average Bond and Galilei numbers ($Bo$=5.0 and $Ga$=59.7) are in good agreement with those of the targeted case \citep[bubble 22 in][with $Bo$= 5.0 and $Ga$= 59.61]{Cano-Lozano2016}. 
\begin{table}[b!]
\centering
\caption{Experimental values of the terminal velocity and terminal Reynolds number obtained for the three cases described in the present work and the different wall distances. All tabulated data refer to the mean value \textpm~standard deviation of all the experiments performed for the cases defined by each row and column.}\label{tab:Re}
\begin{tabular}{c | c | c c c c}
    \hline\hline
    \multicolumn{2}{c}{Case} & $L=1$ & $L=2$ & $L=4$ & $L\rightarrow \infty$\\ [0.5ex]
    \hline
    \multirow{2}{*}{\hspace{1em} 1 \hspace{1em}} & \hspace{2ex}$v$\hspace{2ex} & 1.02\textpm0.04 & 1.07\textpm.04 & 1.06\textpm0.04 & 0.99\textpm0.04\\ 
     & $Re$ & 59.0\textpm0.7 & 62.0\textpm0.6 & 61.6\textpm0.6 & 60.8\textpm0.3\\
    \hline
    \multirow{2}{*}{\hspace{1em} 2 \hspace{1em}} & \hspace{2ex}$v$\hspace{2ex} & 1.06\textpm0.04 & 1.1\textpm0.04 & 1.11\textpm0.04 & 1.15\textpm0.01\\ 
     & $Re$ & 105.3\textpm2.1 & 109.1\textpm0.8 & 109.1\textpm0.4 & 114.6\textpm0.2\\
     \hline
     \multirow{2}{*}{\hspace{1em} 3 \hspace{1em}} & \hspace{2ex}$v$\hspace{2ex} & 0.84\textpm0.01 & 0.83\textpm0.01 & 0.91\textpm0.01 & 0.83\textpm0.02\\ 
     & $Re$ & 90.4\textpm0.6 & 90.0\textpm0.6 & 87.5\textpm0.6 & 99.6\textpm2.7\\
    \hline\hline
 \end{tabular}
 \end{table}
\begin{figure}[t!]
    \centering
    \includegraphics[width=\textwidth]{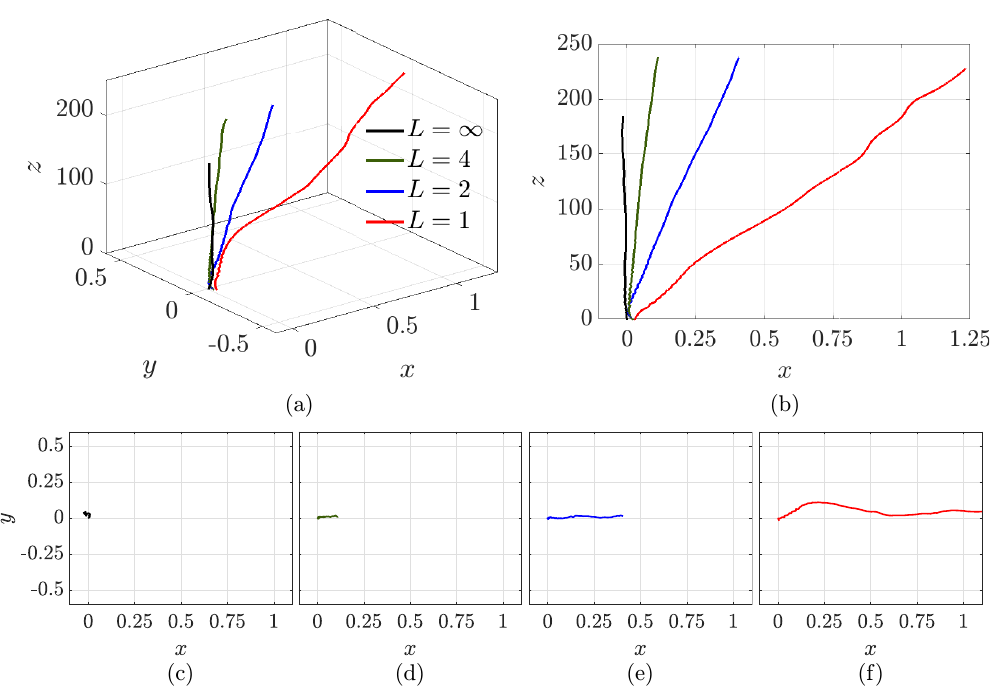}
    \caption{Bubble paths in the rectilinear regime for $L\rightarrow \infty$ (black, $Bo=5.29,~Ga=62.2$), $L$=4 (green, $Bo=4.84,~Ga=58.2$), $L$=2 (blue, $Bo=4.79,~Ga=57.7$), and $L$=1 (red, $Bo=4.77,~Ga=57.6$). (a) Three-dimensional view. (b) Side view ($xz$ plane), perpendicular to the wall plane. Note that horizontal scales in (a) and (b) are magnified with respect to the vertical one to make the migration movement more visible. (c)-(f) Top views (xy plane) for different values of L. The vertical wall, not plotted for the sake of clarity, is placed at plane $yz$ at $x=-4$ for $L=4$, $x=-2$ for $L=2$, and $x=-1$ for $L=1$.}
    \label{fig:trajs_rect}
\end{figure}

Figure~\ref{fig:contours_rect} shows bubble silhouettes at different times in the planes perpendicular and parallel to the wall, $xz$ and $yz$ planes respectively, for three different initial wall distances, together with the unbounded case (Fig.~\ref{fig:contours_rect}a). Note that the bubbles do indeed go up vertically in both planes following a straight path in the case without wall, $L \rightarrow \infty$ (Fig.~\ref{fig:contours_rect}a), in agreement with \citet{Cano-Lozano2016}. In particular, the bubble quickly evolves until reaching a nearly fore-and-aft symmetric oblate ellipsoidal shape (see images depicted in Figs.~\ref{fig:setup}c and \ref{fig:prueba}a). An average major-to-minor bubble diameter aspect ratio $\chi=2.53\pm0.06$ is obtained from all the experiments performed in Case 1. This value agrees fairly well with that obtained using the expression given by \citet{Legendre2012}, $\chi_L=[1-9 We/64 \, (1+0.2 \, Mo^{1/10} \, We)^{-1}]^{-1}=2.19$, 13\% lower than our experimental value, with $We=Bo\,v^2=5.12$, being $Bo=5.22$ and $v=0.99$ (see Table \ref{tab:Re}) the average values corresponding to the unbounded case.  Unlike \citet{Yan2022}, who reported a decrease of $\chi$ in the presence of a vertical wall, no remarkable effects of the wall on the overall bubble aspect ratio have been found in our experiments. Within the rectilinear regime, it is known that the flow around the bubble
exhibits a standing eddy, while it describes a stable straight upward trajectory~\citep{blanco1995structure,Cano-Lozano2013,Cano-Lozano2016}. Although this type of stable bubble has been theoretically and numerically reported~\citep{magnaudet2007wake}, very few experimental works~\citep{zenit2008path} can be found due to the complexity it implies to generate them in a controlled manner, especially for large $Bo$. In this case, when a vertical surface is placed (Figs.~\ref{fig:contours_rect}b-d), the bubble migrates transversely away from the wall as it rises vertically in the plane parallel to the wall, due to the overpressure generated in the liquid film between the bubble and the wall, as can be noticed in both the $xz$ and the $yz$ planes. This effect becomes more evident as the initial wall distance, $L$, decreases, (see Fig.~\ref{fig:contours_rect}d for $L=1$). Consequently, the vertical wall induces a transversal force to the bubble but does not modify the rising regime, keeping a stable rising path regardless of the wall distance. 

To analyze the migration effect in more detail, the dimensionless bubble rising paths have been determined from the time evolution of the bubble centroid (see Fig.~\ref{fig:trajs_rect}). In particular, the three-dimensional paths corresponding to the different initial wall distances, together with the case without the wall, are shown in Fig.~\ref{fig:trajs_rect}(a). In the absence of wall, the bubble indeed follows a nearly vertical route. Considering that the wall is placed in the $yz$ plane, notice that, when the wall is present, the bubble still rises following a straight path, but it exhibits a transversal motion in the $xz$ plane, moving away from the wall, as shown in Figs.~\ref{fig:trajs_rect}(c)-(f). The migration effect is even more evident in Fig.~\ref{fig:trajs_rect}(b), where the paths are plotted in the plane perpendicular to the wall, $xz$. Note that the migration effect increases as the wall is initially closer to the bubble, i.e., for a given height $z$, the bubble is further away as $L$ decreases. In fact, while the maximum transverse motion is almost negligible for $L=4$ ($x\simeq 0.11$), it reaches $x\simeq 1.23$ for $L=1$, as observed in the top views (Figs.~\ref{fig:trajs_rect}c-f). Although for different values of $Bo$ and $Ga$, these results are in general in good agreement with recent numerical simulations performed under similar conditions~\citep{Zhang2020,Yan2022,yan2023three}. It is interesting to point out that the bubble keeps moving away from the wall when it reaches the top of the tank for all the values of $L$, i.e. the migration effect is still present at a height as large as $z\sim$ 200. In particular, contrary to the numerical results for similar cases~\citep{Zhang2020,Yan2022}, where the angle of the rising paths with the vertical plane is shown to decrease as the bubble rises, a nearly constant angle is found in our results. This fact indicates that probably a higher tank would have been required to observe that the wall effect vanishes, being the bubble path vertical again. Nevertheless, as far as we are aware, the vertical distance explored in the present study ($z\gtrsim 200$) is larger than those analyzed experimental and numerically in previous works for similar values of the dimensionless parameters. 

The lateral migration or repulsive force generated by the wall was already reported theoretically~\citep{leal1980particle,Magnaudet2003,sugiyama2010lateral} for small-to-moderate Reynolds number, and explained as an effect of the vorticity diffusion, as well as numerically~\citep{Yang2003,sugioka2015direct,Zhang2020,Yan2022}. These works showed that the flow velocity is restrained in the vicinity of the wall, which breaks the wake symmetry, producing a perpendicular force that causes the lateral motion of the bubble. The migration effect caused by the wall has been also observed in experimental works~\citep{Takemura2002,Takemura2003,Jeong2015,Lee2017}. Nevertheless, the latter are mainly focused on nearly spherical bubbles, corresponding to $Bo$ numbers smaller than those explored in the present work. Specifically, the wall-normal movement agrees with the results obtained by \citet{Takemura2003}, who, for low $Bo$ and low-to-intermediate $Re$ spherical clean bubbles, showed the existence of a transverse force that decreased with the separation distance between the bubble and the wall. They proved that the wall-normal force could be repulsive or attractive to the wall depending on $Re$. Particularly, for $Re<35$, the presence of the wall generated an asymmetric vorticity distribution around the bubble surface, resulting in a transverse force directed away from the wall. For higher $Re$, an acceleration of the flow in the gap between the bubble and the wall was predicted by the irrotational theory, giving rise to a force directed towards the wall. In our results, where the bubbles are ellipsoidal, a repulsive lift is observed, even though $Re>35$, indicating that the dynamics completely change when the bubble deviates from the spherical shape. In this case, the bubble's interface forms a gap with the wall similar to that of a wedge and the liquid moving through generates an overpressure that repels the bubble. Regarding the numerical works, a repulsive lift force is obtained for similar parameters, in good agreement with our results. Specifically, for $L=1$, \citet{Yan2022} calculated a lateral deviation of $x\simeq$ 0.3 at $z$= 90 for a similar $Ga$ but $Bo=2$, while $x\simeq$ 0.5 in our case. Taking into account that $Bo$ is higher in the present work, larger bubble shape deformations take place and the vortex structure becomes more asymmetrical, thus enhancing the wall effect, the agreement is remarkable.  
\begin{figure}[!t]
    \centering
    \includegraphics[width=\textwidth]{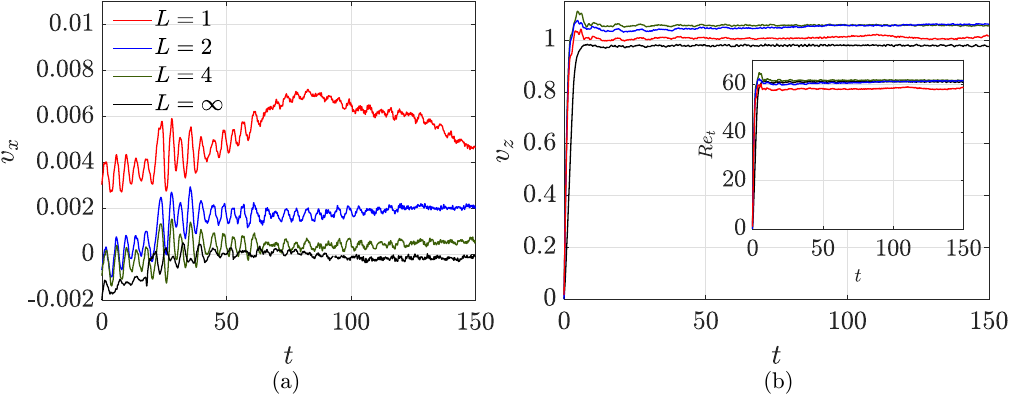}
    \caption{Temporal evolution of (a) the horizontal velocity, $v_{x}$, and (b) the vertical one, $v_z$, for Case 1 (rectilinear) and different wall distances. The inset in panel (b) shows the temporal evolution of the local Reynolds number, $Re_t=v_z\,Ga$.}
    \label{fig:vel_rect}
\end{figure}

The rising bubble motion can be further analyzed by inspecting the bubble velocity, obtained from the time derivative of the bubble paths shown in Fig.~\ref{fig:trajs_rect}. In particular, Fig.~\ref{fig:vel_rect}(a) shows the time evolution of the horizontal velocity perpendicular to the wall, $v_{x}$. As expected, when the bubble rises freely ($L\rightarrow \infty$), the horizontal velocity is nearly negligible. However, when the bubble ascents in the presence of the wall, a positive value of $v_x$ can be clearly observed. For $L> 1$ the horizontal velocity is almost constant during the bubble migration process and increases as $L$ decreases ($v_{x}\approx 0$ for $L\rightarrow \infty$ and $L=4$ and it increases to $v_{x}\approx 0.002$ for $L=2$). Interestingly, for $L=1$ the horizontal velocity initially increases due to the repulsive effect of the wall until it is sufficiently far, when $v_x$ begins to decrease. However, a positive horizontal velocity is still observed when the bubbles reach the top of the tank, indicating that the effect of the wall persists at the final stages of our experiments, even though the bubbles are already far from the wall ($x\simeq 1.2$ at $t\simeq 200$ for $L=1$).
\begin{figure}[t!]
    \centering
    \includegraphics[width=.8\textwidth]{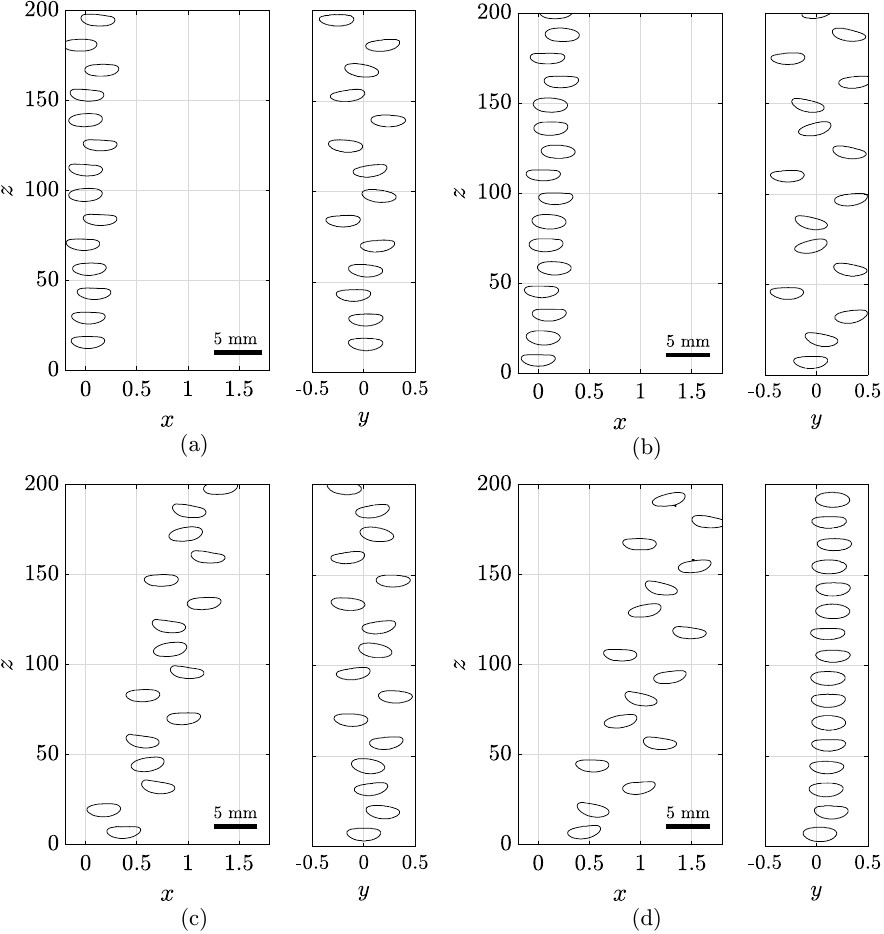}\vspace{-1ex}
    \caption{Bubble silhouettes in the zigzagging regime (Case 2), extracted from the high-speed recordings at different instants of time in the $xz$ (left) and $yz$ (right) planes. (a) $L\rightarrow \infty$, (b) $L=4$, (c) $L=2$, and (d) $L=1$ The vertical wall is placed in plane $xz$ at $x=-4$ ($L=4$), $x=-2$ ($L=2$), $x=-1$ ($L=1$). The bubbles have been scaled down for clarity. Note that the main zigzagging plane does not coincides with plane $xz$ nor $yz$ in cases (a)-(c), as shown in Fig.\ref{fig:trajs_zigzag}.}
    \label{fig:contours_zigzag}
\end{figure}

Figure~\ref{fig:vel_rect}(b) shows the time evolution of the vertical component of the velocity, $v_z(t)$, for the four values of $L$ considered. The vertical velocity rapidly increases from $v_z$= 0 at $t=0$ until it reaches a nearly constant value, i.e., the terminal velocity, $v$. Notice that for all values of $L$, $v\simeq 1$ (see Table~\ref{tab:Re}), indicating that the terminal velocity is practically the gravitational velocity. The small differences observed in the vertical velocity are mostly associated with little differences in the sizes of the bubbles when they are generated rather than due to wall effects. Nevertheless, although a similar evolution is observed, it can be inferred that larger initial accelerations are promoted by the wall, and therefore the terminal velocity is achieved sooner than in the free-rise case. Note also that, after the initial acceleration, the velocity decreases before reaching the terminal velocity. However, when the vertical velocity is expressed in terms of the local Reynolds number, $Re_t(t)=v_z(t)\,Ga$ (inset in Fig.~\ref{fig:vel_rect}b), the Reynolds number associated to the terminal velocity, $Re=v\,Ga$, barely varies with $L$, except for $L=1$, in which a slightly lower value of $Re$ is reached (see Table~\ref{tab:Re}). Although the presence of the wall is expected to reduce $Re$, such an effect is only noticeable when the wall is sufficiently close ($L=1$). The $Re$ reduction by the presence of the wall has been shown to be evident for low $Ga$ numbers but becomes smaller as $Ga$ increases~\citep{Yan2022}. Finally, we would like to remark that the terminal Reynolds number given by \citet{Rastello2011}, valid for free clean bubbles rising within the stable regime, $Re_R=2.05 \, We^{2/3} \, Mo^{-1/5}=61.03$, with $Mo= 9.87 \times 10^{-6}$ and $We= Bo\,v^2=5.12$, agrees very well with the experimental value obtained for $L\rightarrow \infty$, $Re=60.8\pm0.3$. This excellent agreement verifies that bubbles in Case 1 are stable and their surface can be assumed to be clean. Therefore, our results can be considered representative of stable bubbles rising in a pure liquid.

\subsection{Planar zigzag regime}\label{subsec:zig}
Unstable bubbles that rise in a planar zigzag regime were generated using T05 silicon oil and an injector of diameter $d^*$ = 2 mm (Case 2 in Table~\ref{tab:cases}). The resulting average values of the Bond and Galilei numbers ($Bo$=3.87 and $Ga$=98.7) reasonably agree with those of bubble 19 in \citet{Cano-Lozano2016}, with $Bo$=4.0 and $Ga$=100.8.
\begin{figure}[!t]
    \centering
    \includegraphics[width=\textwidth]{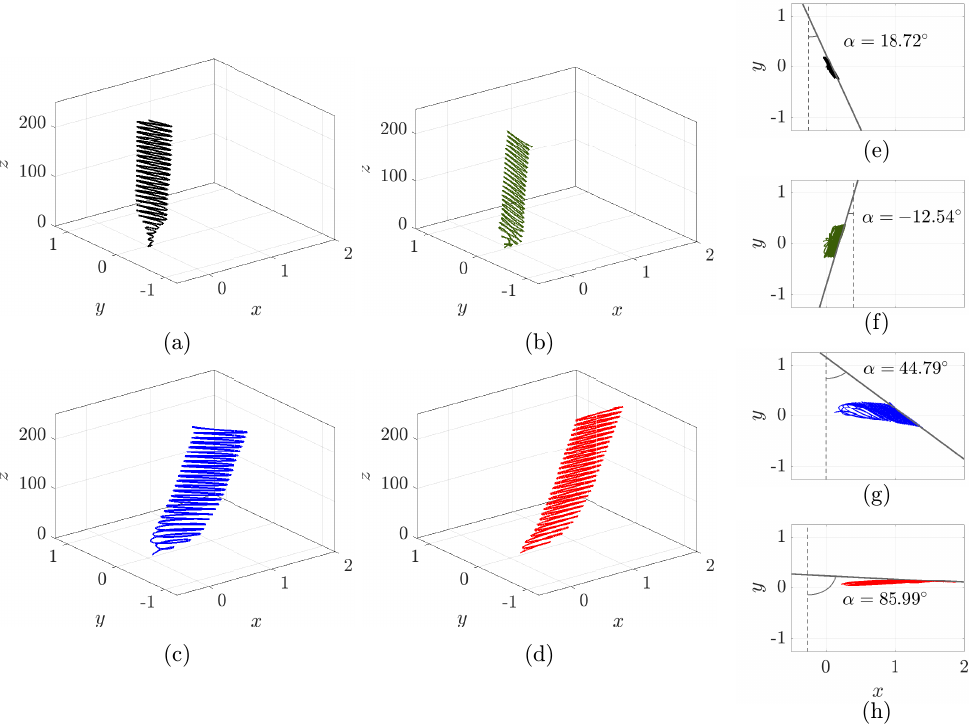}   
    \caption{(a)-(c) Three-dimensional rising path corresponding to the planar zigzagging regime (Case 2) for (a) $L\rightarrow \infty$ ($Bo=3.86,~Ga=98.6$) (b) $L=4$ ($Bo=3.85,~Ga=98.3$) (c) $L=2$ ($Bo=3.87,~Ga=98.7$) and (d) $L=1$ ($Bo=3.89,~Ga=99.2$). Corresponding top views for (e) $L\rightarrow \infty$ (f) $L=4$ (g) $L=2$ and (h) $L=1$, where the angle that the main zigzagging plane forms with the wall, $\alpha$, is indicated. The vertical wall, not plotted for the sake of clarity, is placed at plane $yz$ at $x=-4$ for $L=4$, $x=-2$ for $L=2$, and $x=-1$ for $L=1$.}
    \label{fig:trajs_zigzag}
\end{figure}

Figure~\ref{fig:contours_zigzag} shows the silhouette of an individual bubble at different times as it rises, in planes perpendicular ($xz$) and parallel ($yz$) to the wall for $L\rightarrow \infty$, 4, 2, and 1. Note that in this case, the bubble does not rise with a stable path, but oscillations in both planes $xz$ and $yz$ are observed. In fact, oscillations are mainly contained in a vertical plane, as shown in Fig.~\ref{fig:trajs_zigzag}, described in detail below, indicating that this case corresponds to a planar zigzagging regime. In this case, the ellipsoidal shape of the bubble loses its fore-and-aft symmetry (see Fig.~\ref{fig:contours_zigzag} and the central image of Fig.~\ref{fig:setup}c), in agreement with previous experimental~\citep{brucker1999structure,zenit2008path,zenit2009measurements} and numerical results~\citep{mougin2002paht,mougin2006wake,Cano-Lozano2016} on freely rising bubbles. 

An average aspect ratio $\chi$= 2.57 $\pm$ 0.09 is obtained, in reasonable agreement with the experimental results by \citet{zenit2008path} for the same oil. According to \citet{Legendre2012}, considering the values corresponding to the free rise ($We=Bo \,v^2=$ 5.13 and $Mo=0.61 \times 10^{-6}$), $\chi_L=[1-9 We/64 \, (1+0.2 \, Mo^{1/10} \, We)^{-1}]^{-1}=2.38$, being 7.4\% lower than our experimental value. Note that in this case, the average value of the Bond number is similar to that of Case 1 (Table \ref{tab:cases}), although the viscosity of the liquid used (T05 oil) is lower than that of the liquid used in the rectilinear case (T11 oil). This increases the corresponding $Ga$ (and consequently the Reynolds number) and the bubble paths become unstable. Bubble shape and path instability are related to the wake structure behind the bubble. Particularly, the three-dimensional wake consists of two counter-rotating trailing vortices with a symmetry plane, that change their sign twice during the zigzag period~\citep{Cano-Lozano2016}.
\begin{figure}[t!]
    \centering
    \includegraphics[width=\textwidth]{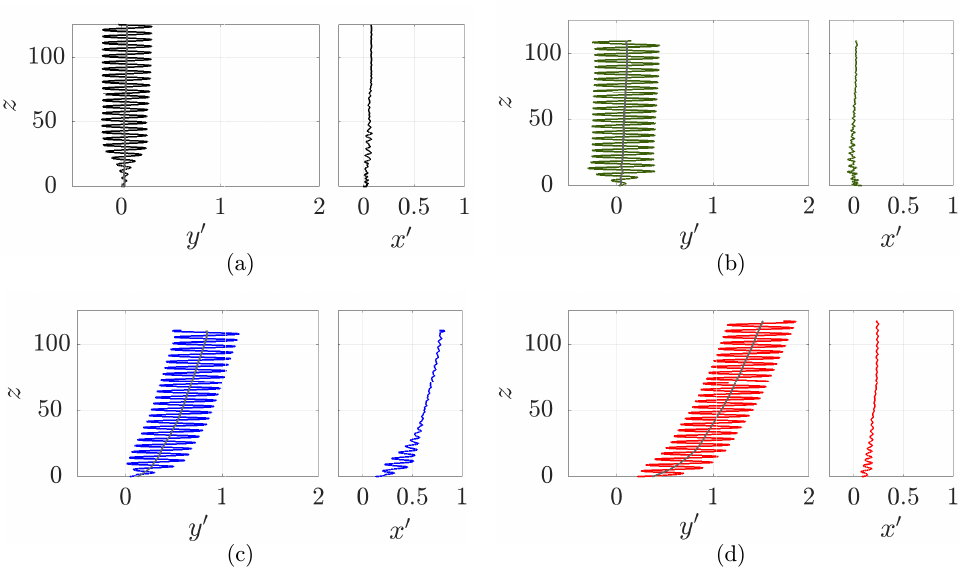}   
    \caption{Rising trajectories in the main zigzagging plane, $y'z$, (left) and the orthogonal plane, $x'z$, (right) corresponding to the cases in Fig. \ref{fig:trajs_zigzag} for (a) $L\rightarrow \infty$ ($\alpha=18.72^{\circ}$), (b) $L=4$ ($\alpha=-12.54^{\circ}$), (c) $L=2$ ($\alpha=44.79^{\circ}$), and (d) $L=1$ ($\alpha=85.99^{\circ}$). Central grey lines depict the average path.}
    \label{fig:xz_yz_zigzag}
\end{figure}

The three-dimensional rising paths corresponding to individual experiments have been reconstructed from the bubble centroid position detected in both $xz$ and $yz$ planes, and are plotted in Figs.~\ref{fig:trajs_zigzag}~(a)-(d). In the absence of wall (Fig.~\ref{fig:trajs_zigzag}a), it is shown that, after some transient oscillations, the bubble begins to oscillate in a plane that, in this particular case, forms an angle $\alpha=18.72^{\circ}$ with the $yz$ plane. Furthermore, when the wall is present (Figs.~\ref{fig:trajs_zigzag}b-d), the bubbles still follow a planar zigzagging trajectory and no change is observed in the rising regime. However, as in the rectilinear regime, the wall causes the bubble to move away from it as the bubble rises, being this effect more pronounced as $L$ decreases, in agreement with recent numerical studies~\citep{Zhang2020,yan2023three}. Indeed, when the bubble rises near a wall, ~\citet{yan2023three} showed numerically that the vortex closest to the wall gets smaller, giving rise to an asymmetric vortex structure that causes the bubble to migrate away from the wall. These characteristics can be better analyzed from the corresponding top views shown in Figs.~\ref{fig:trajs_zigzag}(e)-(h), where the bubble oscillates in planes that move away from the wall in each cycle (especially evident in Fig.~\ref{fig:trajs_zigzag}g). Nevertheless, the angle $\alpha$ that the oscillation plane forms with the wall, or similarly with plane $yz$, varies in each case. In fact, from our experiments, it has been observed that when the wall is far from the bubble, $\alpha$ is established randomly and it is not controllable. That is, when the bubble rises freely or sufficiently far from the wall ($L=2$, $L=4$), each individual experiment performed with the same wall distance exhibits a distinct zigzagging plane orientation, being different from that in cases plotted in Fig.~\ref{fig:trajs_zigzag}. This result, which agrees with \citet{Cano-Lozano2016}, indicates that the experimental facility does not introduce any perturbation triggering the instability and imposing a preferential zigzagging plane. However, when $L$ is sufficiently small the bubble trajectory tends to oscillate in a plane perpendicular to the wall (see Fig.~\ref{fig:trajs_zigzag}h). This phenomenon has been reported in numerical studies~\cite{Zhang2020}, but, as far as we are aware, this is the first experimental proof of such a phenomenon. Interestingly, the zigzagging regime is established earlier when the wall is present since it imposes a lateral disturbance triggering the instability. Then, the presence of the wall is observed to promote the onset of the oscillations in a normal direction. This fact can be attributed to the interaction of the bubble wake with the wall, as suggested in numerical works in which the liquid flow field is analyzed~\citep{Yan2022}. In particular, the wall is shown to modify the flow field around the bubble, the vortex structure, and the bubble shape, which becomes asymmetric along the central axis. Specifically, the bubble presents different curvatures at the near-wall and the free sides, what causes a certain vorticity accumulation on the bubble surface that imposes the shedding of periodic vortices. The latter are observed to be symmetrical in the plane parallel to the wall, while they lose their symmetry in the wall-normal plane, imposing a wall-normal zigagging plane together with the migration effect.
\begin{figure}[t!]
    \centering
    \includegraphics[width=0.9\textwidth]{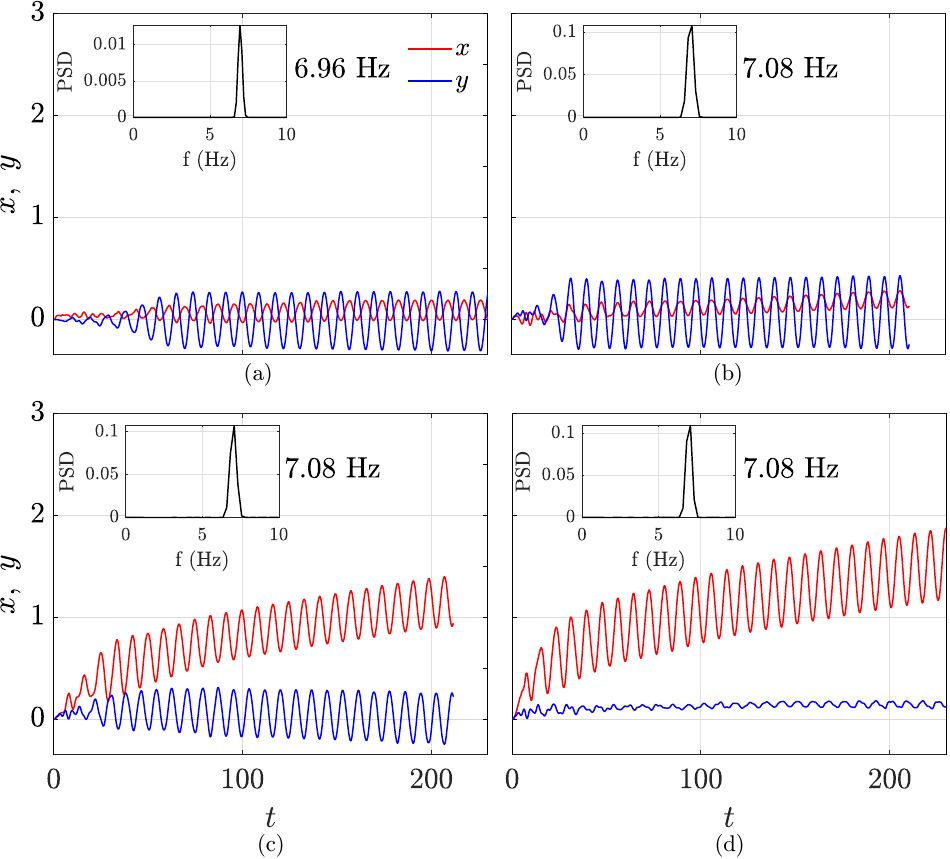}
    \caption{Time evolution of the horizontal coordinates of the bubble trajectory, namely, $x(t)$ (red line) and $y$ (blue line), corresponding to cases in Fig.\ref{fig:trajs_zigzag}, for (a) $L\rightarrow \infty$ ($\alpha=18.72^{\circ}$), (b) $L=4$ ($\alpha=-12.54^{\circ}$), (c) $L=2$ ($\alpha=44.79^{\circ}$), and (d) $L=1$ ($\alpha=85.99^{\circ}$). Insets in each subfigure show the PSD of $x(t)$, exhibiting a clear peak at $f\simeq7.08$ Hz in the bounded cases.}\label{fig:xy_vs_t_PSD_zigzag}
\end{figure}

To directly compare the zigzag paths for different values of $L$, taking into account the value of $\alpha$ in each case, paths are plotted in the main zigzagging plane $y'z$, and its orthogonal view, $x'z$ in Fig.~\ref{fig:xz_yz_zigzag}, being $y'=x \sin{\alpha}-y \cos{\alpha}$ and $x'= x \cos{\alpha}+y \sin{\alpha}$. In this way, it can be confirmed that while the oscillations take place mainly in plane $y'z$, they are almost negligible in the perpendicular one, $x'z$. This corroborates that a planar zigzagging regime is indeed established. Additionally, the migration effect caused by the wall is more evident as $L$ decreases, as observed in the mean path depicted in each case. Regarding the zigzagging amplitude, it slightly increases as the bubble rises in all cases. This indicates that the path is still not completely developed at $z\simeq200$, differently from previous works on free bubble rise, where the regime is fully established for shorter heights~\citep{Cano-Lozano2016}. Moreover, smaller vertical distances have been used in previous numerical works including a vertical wall for similar regimes~\citep{Zhang2020,Yan2022,yan2023three}, in which fully developed rising paths are obtained for shorter vertical distances. Concerning the effect of the wall on the oscillating amplitude, it slightly increases when the wall is present, being $A \simeq \pm 0.25$ for $L\rightarrow \infty$, while $A\simeq \pm$0.3, $\pm$0.32 and $\pm$0.34 for $L$=4, 2 and 1, respectively. The increase of the amplitude in the bounded cases seems to be due to the destabilizing effect introduced by the wall~\citep{Zhang2020}, which also explains the slight increase of $A$ as $L$ decreases. Nevertheless, the quantitative effect of $L$ on the amplitude is hard to infer since the angle $\alpha$ of the zigzagging plane is different in each case and may vary with time.

To further analyze the bubble movement in this regime, Fig.~\ref{fig:xy_vs_t_PSD_zigzag} shows the time evolution of the path coordinates, $x(t)$ and $y(t)$. The migration effect of the wall on the bubble trajectory is clear. While $y(t)$ oscillates around $y=0$ in all cases, when the wall is present $x(t)$ oscillates around a value that increases with time because the bubble migrates away from the wall. Again, the migration effect is more evident as $L$ decreases. Furthermore, note that for $L=1$ (Fig.~\ref{fig:xy_vs_t_PSD_zigzag}d), $y(t)$ barely changes with time since the oscillating plane is nearly perpendicular to the wall. In general, once the zigzagging regime is achieved, a constant oscillation frequency is established, whose value is not affected by the wall, as can be clearly observed in the power spectral density graph shown as insets in each panel of Fig. \ref{fig:xy_vs_t_PSD_zigzag}. In particular, a frequency $f\simeq$ 7.08 Hz is obtained for the bounded cases depicted. Taking all the experiments carried out within this regime, the mean oscillation frequency is $f=7.10 \pm 0.10$ Hz, which provides a Strouhal number $St=f \, D^*/v^* =0.109 \pm 0.004$, as displayed in Table~\ref{tab:cases}. An excellent agreement is achieved with the numerical result $St=0.108$, obtained for this particular bubble in \citet{Cano-Lozano2016} for the unbounded case. The phase shift between $x(t)$ and $y(t)$ depends on the angle $\alpha$, being in phase when $\alpha<0$ (Fig.~\ref{fig:xy_vs_t_PSD_zigzag}b) and out of phase for $\alpha>0$ (Figs.~\ref{fig:xy_vs_t_PSD_zigzag}a,c).
\begin{figure}[!t]
    \centering
    \includegraphics[width=\textwidth]{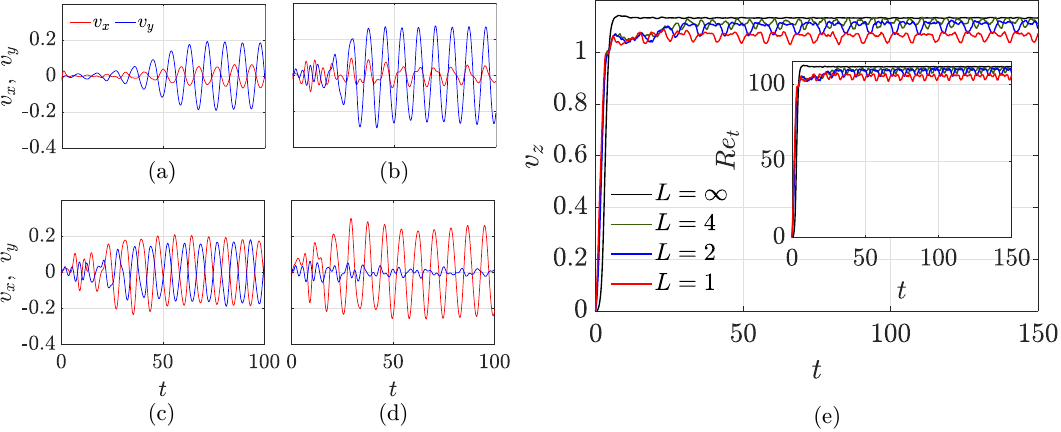}
    \caption{(a)-(d) Time history of the velocity horizontal components, $v_x$, $v_y$ for a) $L\rightarrow \infty$ ($\alpha=18.72^{\circ}$), (b) $L=4$ ($\alpha=-12.54^{\circ}$), (c) $L=2$ ($\alpha=44.79^{\circ}$), and (d) $L=1$ ($\alpha=85.99^{\circ}$). Velocities for $t\leq$100 are shown for the sake of clarity. (e) Vertical velocity $v_z$ as a function of time for the four wall distances. The inset shows the time evolution of the local Reynolds number.}
    \label{fig:vh_vs_t_PSD_zigzag}
\end{figure}

The bubble velocity components extracted from the trajectories are displayed in Fig.~\ref{fig:vh_vs_t_PSD_zigzag}. As expected, the horizontal components of the velocity begin to oscillate after a transient time that decreases as $L$ is reduced, indicating that the wall favors the onset of the instability (Figs. ~\ref{fig:vh_vs_t_PSD_zigzag}a-d). The velocity components exhibit positive and negative values associated with the bubble displacements. As in the trajectories, the oscillation amplitude of $v_x$ and $v_y$, as well as the phase difference, are related to the lateral displacement and depend on the angle $\alpha$. The mean migration velocity of the bubbles when the wall is present, of order $\mathcal{O}(10^{-3})$, is much smaller than the oscillation amplitude, of order $\mathcal{O}(10^{-1})$, and cannot be observed in the plots. With respect to the vertical velocity, Fig.~\ref{fig:vh_vs_t_PSD_zigzag}(e) shows that, as in the rectilinear regime, the rise velocity initially increases linearly with time until it reaches a state where it oscillates around a mean value, slightly larger than the gravitational velocity in this case. In fact, in this regime, given that the values of $Ga$ are larger than the critical one corresponding to our experimental mean value of $Bo=$ 3.9, the occurring path instability makes $v_z(t)$ oscillate around the mean value, $v$, at a frequency twice that of the zigzagging motion, $f\simeq 14$ Hz, as anticipated numerically by \citet{Cano-Lozano2016} for a free-rising bubble. The same temporal evolution can be observed for the local Reynolds number depicted as an inset. The mean values of $v$, as well as those of associated $Re$ are shown in Table \ref{tab:Re}. The mean terminal velocity (and thus the Reynolds number) slightly decreases as $L$ decreases, due to the additional drag induced by the presence of the wall. Based on the experiments by \citet{Maxworthy1996}, \citet{Cano-Lozano2015} proposed a correlation to predict the Reynolds number for clean unstable bubbles, providing $Re_{CL}=[We^2\, (We-2.14)/0.505 \, Mo]^{1/4}=126.4$, with $We=Bo \, v^2$=5.13, $Bo$=3.88 and $v$=1.15 the values corresponding to the free case. Note that $Re_{CL}$ is only 10\% larger than $Re=114.6$, the average experimental value for $L\rightarrow \infty$ (Table \ref{tab:Re}), indicating that bubbles in Case 2 can be considered to be clean.

\subsection{Spiral regime}\label{subsec:spi}
 \begin{figure}[t!]
    \centering
    \includegraphics[width=.8\textwidth]{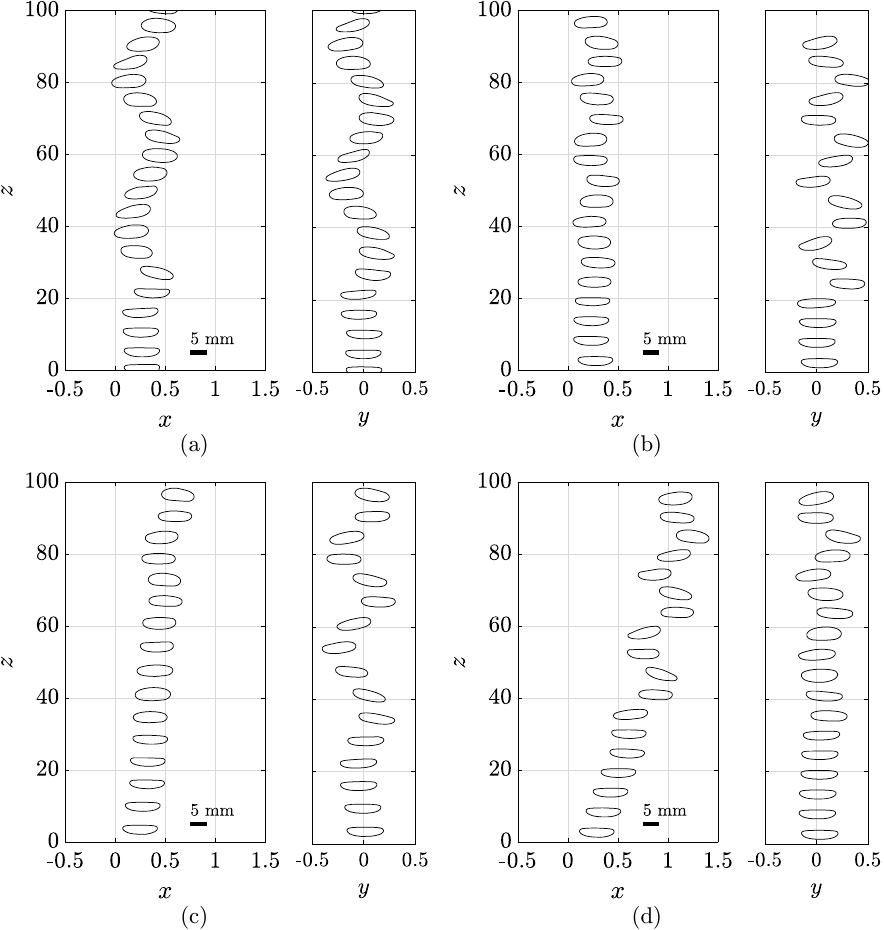}\vspace{-2ex}
    \caption{Bubble silhouettes in the spiral regime (Case 3), extracted from the high-speed recordings at different instants of time in the $xz$ (left) and $yz$ (right) planes. (a) $L\rightarrow \infty$, (b) $L=4$, (c) $L=2$, and (d) $L=1$. The bubbles have been scaled down for clarity. The vertical wall, not plotted for the sake of clarity, is placed at $x=-4$ for $L=4$, $x=-2$ for $L=2$, and $x=-1$ for $L=1$.}
    \label{fig:contours_spiral}
\end{figure}
Unstable bubbles that rise in a spiral path were generated using glycerol-water mixtures (74.16\%  - 74.89\% weight) and a 12 mm superhydrophobic circular substrate (Case 3 in Table~\ref{tab:cases}). The resulting average values of the Bond, $Bo=10.29$, and Galilei, $Ga=108$, numbers are similar to those of the target case \citep[bubble 26 in][with $Bo$=10 and $Ga$=100.25]{Cano-Lozano2016}. Experiments in this case were not as reproducible as those in Cases 1 and 2, since the spiral regime takes longer to develop. Moreover, the physical properties of GW mixtures are not as stable as silicon oils, and therefore the values of the control parameters slightly varied from experiment to experiment. In addition, the established rising paths depended on the initial injection conditions, which were not exactly the same in all the experiments, since the bubbles were generated using a superhydrophobic substrate (see \citet{Rubio-Rubio2021} for more details) instead of a conventional injector as in Cases 1 and 2. Furthermore, bubble shape oscillations were observed along the rise of these bubbles. However, the deformation of the bubble never led to a collision with the wall even for the smallest values of $L$.
\begin{figure}[t!]
    \centering
    \includegraphics[width=\textwidth]{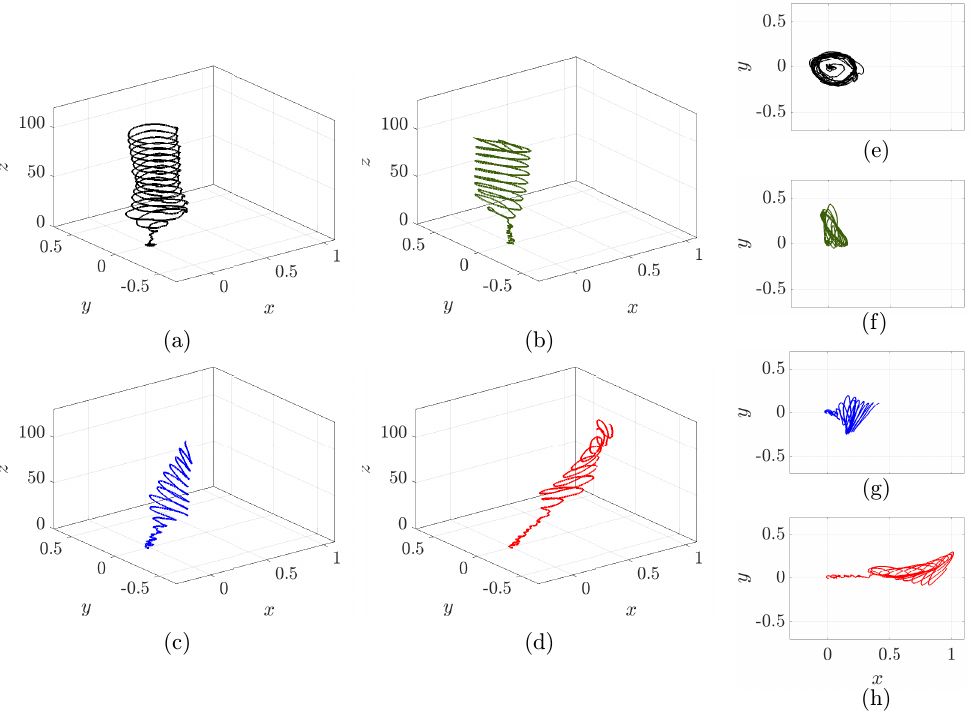}     
    \caption{(a)-(c) Three-dimensional rising path corresponding to the spiral regime (Case 3) for (a) $L\rightarrow \infty$ ($Bo=9.94,~Ga=121.4$) (b) $L=4$ ($Bo=10.37,~Ga=103.0$) (c) $L=2$ ($Bo=10.38,~Ga=103.1$) and (d) $L=1$ ($Bo=10.43,~Ga=103.5$). Corresponding top views for (e) $L\rightarrow \infty$ (f) $L=4$ (g) $L=2$ and (h) $L=1$. The vertical wall, not plotted for the sake of clarity, is placed at $x=-4$ for $L=4$, $x=-2$ for $L=2$, and $x=-1$ for $L=1$.}
    \label{fig:trajs_spiral}
\end{figure}

Figure~\ref{fig:contours_spiral} shows the bubble silhouettes at different times in both transverse ($xz$) and parallel ($yz$) planes for three different initial wall distances and for the unbounded case. Like in the previous cases, the bubble silhouettes have been scaled down so they don’t look so big in the figures. Note that oscillations in both planes take place and, as observed in Fig.~\ref{fig:trajs_spiral}, the bubbles rise describing an elliptical spiral or a zigzag trajectory, depending on the wall distance. In this case, as seen in Fig.~\ref{fig:setup}(c), the bubble deformation is stronger than in Case 2 (zigzag), presenting large deviations from ellipsoids. This large deformation is attributed to the reduced effect of surface tension since the Bond and the Weber numbers are higher than in the previous cases (see Table~\ref{tab:Re}). In particular, besides fore-and-aft asymmetry, right-left asymmetry along the major diameter can also be observed, presenting a more pointed rim at the exterior part of the path, in agreement with the numerical shapes obtained by \citet{Cano-Lozano2016} for the same bubble and by \citet{Zhang2020} and \citet{yan2023three} for cases under similar conditions. The latter works show that the wake structure of this regime consists of a double spiral composed of double-threaded vortex pairs. Specifically, an average aspect ratio $\chi$= 2.97 $\pm$ 0.17 is obtained, in reasonable agreement with the experimental results by \citet{zenit2008path} using a silicon oil of the same Morton number. According to the expression given by \citet{Legendre2012}, $\chi_L=[1-9 We/64 \, (1+0.2 \, Mo^{1/10} \, We)^{-1}]^{-1}=3.27$, with $Mo=5.0 \times 10^{-6}$ and $We=Bo \,v^2=$6.96 the mean value provided with $v=0.83$ (Table \ref{tab:Re}) and $Bo=10.11$ (values for $Mo$, $Bo$ and $v$ correspond to the unbounded cases, $L\rightarrow \infty$). Notice that a good agreement is achieved, being $\chi_L$ 10\% larger than our experimental value.

The rising paths are shown in Fig.~\ref{fig:trajs_spiral} where the freely ascending bubble first exhibits a flattened spiraling trajectory that eventually evolves into a nearly circular helix (Fig.\ref{fig:trajs_spiral} a,e). The dimensionless vertical distances reported in this case are smaller than in Cases 1 and 2 because the bubble diameters are larger (see Table \ref{tab:cases}). As in the rectilinear and the zigzagging cases, an overall migration effect is also observed when the wall is present (Fig.\ref{fig:trajs_spiral} b-d). This effect, similar to the other regimes, is promoted by the asymmetric vortex structures generated in the wake of the bubbles when the wall is present. However, unlike in the other cases, a change in the rising regime is observed here. Although a direct comparison with the free case must be done with caution since $Ga$ is lower in the wall-bounded cases, it is clearly seen that the spiral regime is no longer established when the wall is present (Fig.\ref{fig:trajs_spiral} f-h). Instead, a flattened spiral, or even almost a zigzag pattern, is established. Since the wall is already present when the bubble is released, it imposes a boundary condition that is different from that given in the free-rising case, preventing the bubble from following a complete spiral path from the very beginning. Such spiraling or zigzag trajectory is expected to be established at sufficiently long distances from the injection point when the bubble is already far from the wall. For example, for $L=$ 4, the bubble describes an elliptic spiraling path whose major axis is tilted with respect to the wall while the $L=2$ case is prone to start following a zigzagging regime in a plane nearly parallel to the wall and develops a flattened spiral path as it evolves. However, for $L=1$ the bubble describes a flattened spiraling path whose major and minor axes rotate as the bubble rises. This behavior differs from recent numerical works~\citep{Zhang2020}, where only the migration effect is observed when the wall is present, but the bubble keeps its helical trajectory. 
\begin{figure}[t!]
    \centering
    \includegraphics[width=\textwidth]{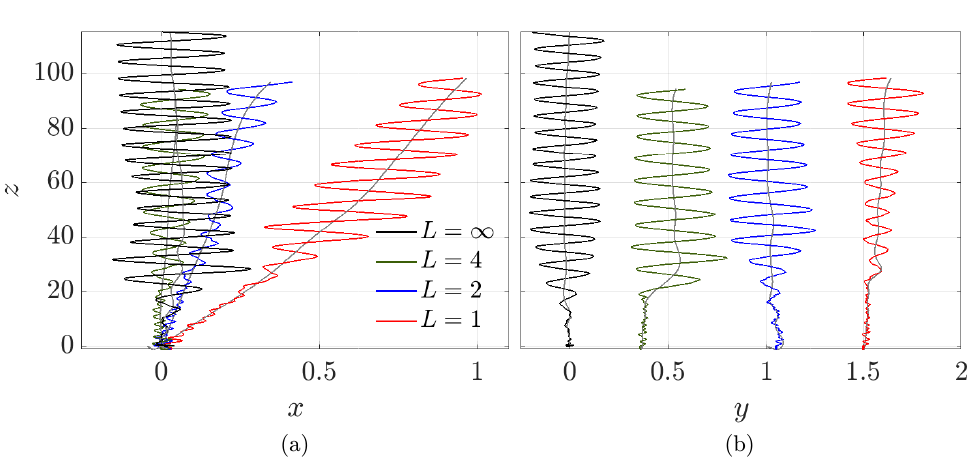}
    \caption{(a) Rising paths corresponding to the spiraling regime for the different values of $L$ in the (a) wall-normal plane ($xz$) and (b) wall-parallel plane ($yz$). In (b), the centers of the trajectories were displaced $\sim$ 0.5 units each in the positive $y$ direction for the sake of clarity. Grey lines denote the mean rising path. }\label{fig:xz_spiral} 
\end{figure} 

\begin{figure}[t]
    \centering
    \includegraphics[width=0.85\textwidth]{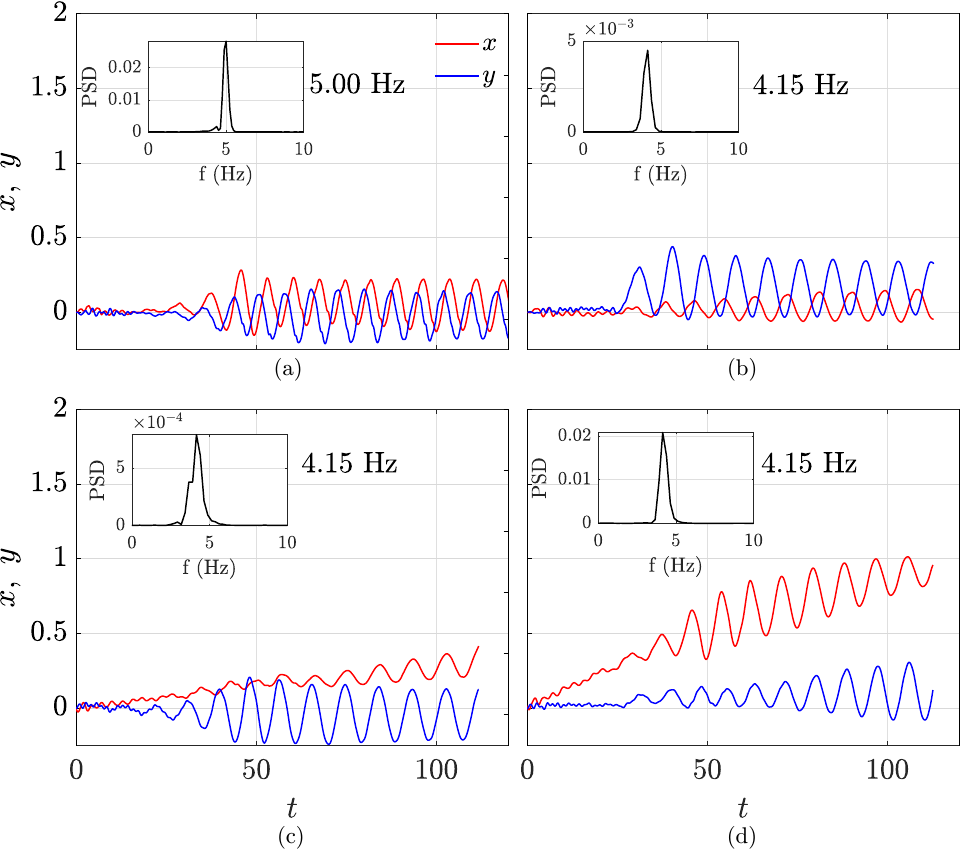}
    \caption{Time evolution of horizontal coordinates of the bubble centroid, namely, $x(t)$ (red line), $y$ (blue line) corresponding to cases in Fig.\ref{fig:trajs_spiral}}, for a) $L\rightarrow \infty$, (b) $L=4$, (c) $L=2$, and (d) $L=1$. Insets in each subfigure show the PSD of $x(t)$.\label{fig:xy_vs_t_PSD_spiral}
\end{figure}

The migration effect caused by the wall can be clearly seen in Fig.~\ref{fig:xz_spiral}(a), where the projection of the bubble rising path on the plane perpendicular to the wall is shown. As in Cases 1 and 2, the average bubble trajectory is nearly vertical in the absence of the wall, and it moves further away from the wall as $L$ decreases. However, no wall effect is observed in the projection of the mean trajectory of the bubble in the plane parallel to the wall, $yz$, which, after a transient period in which it deviates from the injection point, follows an almost vertical trend (Fig.~\ref{fig:xz_spiral}b). Unlike Case 2, paths exhibit oscillations in both $xz$ and $yz$ planes since they are three-dimensional in this regime. Note that, for $L=1$, the amplitude of the oscillations increases until $z\approx 55$ in the $xz$ plane and begins to decrease for $z> 55$, while it keeps increasing in the $yz$ plane, indicating that the major axis of the elliptic spiraling path is rotating. Continuing with the oscillation amplitude, shown in Fig.~\ref{fig:xy_vs_t_PSD_spiral} as a function of time, while the amplitudes are comparable in both planes for $L\rightarrow \infty$ because a nearly helical path is established, $L=4$ and $L=2$ cases show much larger amplitudes in one of the planes since they follow flattened spiral paths as mentioned above. The case in which $L=1$ is more complex, where a flattened spiral path is developing as the bubble rises and migrates away from the wall. A more detailed analysis of the effect of the wall on the trajectory amplitude cannot be performed because the rising regimes are not exactly the same for each $L$. Considering the oscillation frequency (insets in Figs.~\ref{fig:xy_vs_t_PSD_spiral} a-d), a constant value ($f=4.15$ Hz) slightly lower than that of the free case ($f=5.00$ Hz) has been found independently of $L$. The higher oscillating frequency in the unbounded case, whose Strouhal number is $St=0.160$, can be attributed to the fact that the corresponding Galilei number is slightly higher than those of the wall-bounded cases, where $St=0.132$. Nevertheless, these values are in good agreement with those reported by~\citet{Cano-Lozano2016} for bubble 26, who found $St$ to vary from 0.136 to 0.174 as the bubble evolves from a planar zigzag to a spiral path.\\
\begin{figure}[t]
    \centering
    \includegraphics[width=0.95\textwidth]{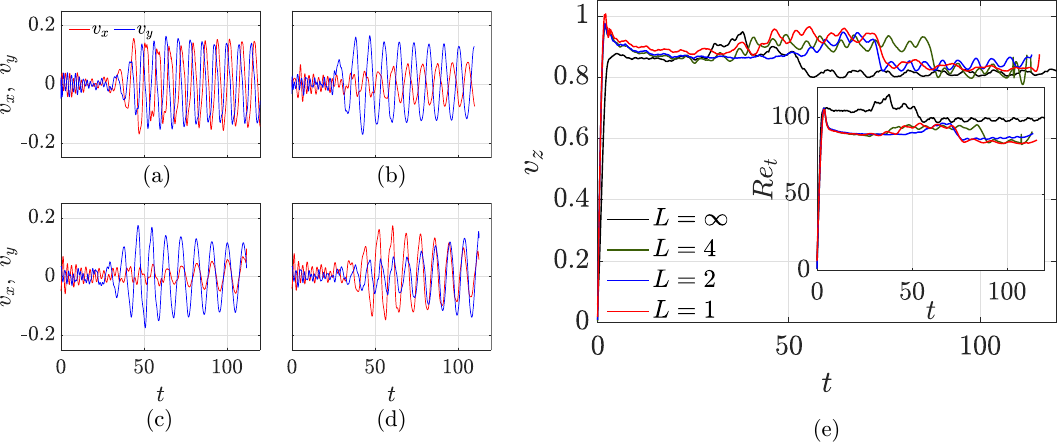}
    \caption{Time history of the velocity horizontal components, $v_x(t)$ and $v_y(t)$ corresponding to cases in Fig.~\ref{fig:trajs_spiral}. (a) $L\rightarrow \infty$, (b) $L=4$, (c) $L=2$, and (d) $L=1$. (e) Vertical velocity $v_z(t)$ as a function of time for all the different wall distances. The inset represents the time evolution of the local Reynolds number.}
    \label{fig:vh_vs_t_PSD_spiral}
\end{figure}

The velocity components extracted from the trajectories are presented in Fig.~\ref{fig:vh_vs_t_PSD_spiral}, where the horizontal components (Fig.~\ref{fig:vh_vs_t_PSD_spiral} a-d) can be observed to oscillate with time at the same frequency as the corresponding trajectories (Fig.~\ref{fig:xy_vs_t_PSD_spiral} a-d). Similarly to Case 2, both components $v_x(t)$ and $v_y(t)$ present positive and negative values as the bubble alternatively approaches and moves away from the wall. The oscillation amplitude of $v_x$ and $v_y$, as well as the phase difference, are related to the orbital excursions described above. Notice that the amplitude of the velocity oscillations decreases when the wall is present since different regimes are established, i.e. a helical regime for $L\rightarrow \infty$ and tends to develop a flattened spiral one as $L=$ decreases. Regarding the vertical velocity, in this case, it increases, reaching a quasi-terminal value before dropping and beginning to oscillate around a mean value smaller than the gravitational velocity with an amplitude that decreases with the presence of a wall (Fig.~\ref{fig:vh_vs_t_PSD_spiral}e). The same temporal evolution can be observed for the local Reynolds number depicted as an inset. The drop in the velocity (Reynolds) may be attributed to the increase in drag when the spiral regime is established, which is retarded by the presence of the wall. Note that $Re_t$ decreases when the wall is present, a feature already observed in the rectilinear and zigzagging regimes, especially for $L=1$. The mean values of the terminal velocity $v$, as well as those of $Re$ are shown in Table \ref{tab:Re}. The Reynolds number predicted by the correlation by \citet{Cano-Lozano2015} gives $Re_{CL}=[We^2(We-2.14)/0.505Mo]^{1/4}=98.12$, which compares very well with the average experimental value for $L\rightarrow \infty$, $Re=99.6$ (Table \ref{tab:Re}). This excellent agreement indicates that contamination, although probably present due to the use of water, did not alter the experimental results obtained here. 

\subsection {Comparison of regimes} \label{subsec:comp}
\begin{figure}[t!]
    \centering
    \includegraphics[width=1\textwidth]{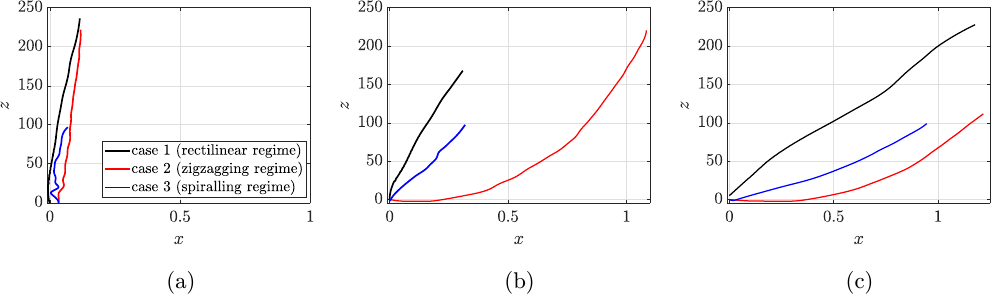}
    \caption{Rising path in $xz$ plane for the rectilinear regime (black), together with the average path for zigzag (red) and spiral (blue) regimes corresponding to (a) $L=4$, (b) $L=2$, and (c) $L=1$.}\label{fig:xz_comparison}
\end{figure}

In this section, the effect of the bubble-rising regime on the migration process is analyzed. To that aim, the average rising paths in the vertical $xz$ plane, normal to the wall, for the different regimes are directly compared for a given value of $L$. In particular, the rectilinear path in the $xz$ plane is displayed in Fig.~\ref{fig:xz_comparison} together with the corresponding averaged paths of zigzagging and spiraling cases for $L=4$ (Fig.~\ref{fig:xz_comparison}a), $L=2$ (Fig.~\ref{fig:xz_comparison}b) and $L=1$ (Fig.~\ref{fig:xz_comparison}c). For both unstable regimes, the depicted paths correspond to those shown in solid gray lines in Figs.~\ref{fig:xz_yz_zigzag} and \ref{fig:xz_spiral}. It can be observed that, for the three values of $L$, the zigzag regime (Case 2) shows the largest average lateral displacement away from the wall, while the rectilinear regime (Case 1) is the least affected. The same result is presented for all the analyzed values of $L$, being more evident as $L$ decreases. Thus, the zigzag regime, besides presenting the largest amplitude, as shown above, also promotes the strongest bubble lateral migration away from the wall. This result is related to the characteristic wake structure that the bubble presents when rising within this regime, which has been described above. Notice that, when comparing the three regimes, both $Bo$ and $Ga$ vary. On the one hand, cases 1 and 2 present a similar $Bo$, while $Ga$ is larger for Case 2. On the other hand, $Ga$ is comparable in cases 2 and 3, being $Bo$ larger for Case 3. These results represent relevant findings since they can be used to optimize many applications involving bubble-wall interaction phenomena. 

\subsection{Effect of the vertical position of the wall leading edge}\label{subsec:wall}
\begin{figure}[b!]
    \centering
    \includegraphics[width=1\textwidth]{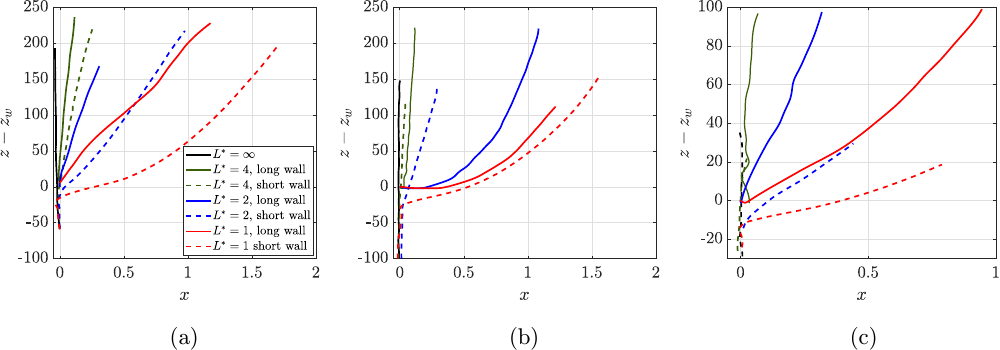}
    \caption{Bubble paths in the lateral plane for (a) rectilinear, (b) zigzagging, (c) spiraling regimes. Solid and dashed lines correspond to a long wall and a short wall, respectively. Here, (b) and (c) show averaged paths.}\label{fig:xz_short_wall}
\end{figure}
Additional experiments were carried out in order to investigate the effect of the vertical distance from the air injector to the wall edge. To that aim, unlike in the initial experiments, a short wall was placed into the tank. In this case, the wall edge did not reach the bottom of the tank and the bubbles initially began to rise freely before reaching the wall. The vertical distance from the wall edge to the injector, $z_w$, was fixed so that the bubbles reached the terminal velocity in each regime before they encountered the wall. In particular, $z_{w}\approx60$ for the rectilinear regime, $z_{w}\approx160$ for the zigzagging regime, and $z_{w}\approx85$ for the spiraling one.
\begin{figure}[t!]
    \centering
    \includegraphics[width=1\textwidth]{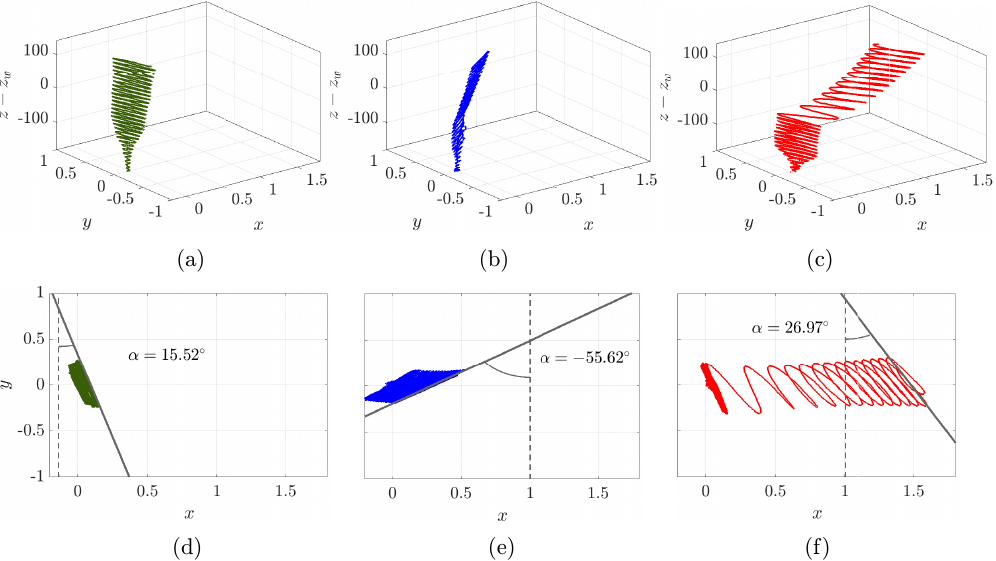}
    \caption{3D (top row) and xy (bottom row) bubble paths for zigzagging regime using a short wall and (a), (d) $L$=4 (green, $Bo=3.27,~Ga=86.4$), (b), (e) $L$=2 (blue, $Bo=3.28,~Ga=87.3$), and (c), (f) $L$=1 (red, $Bo=3.24,~Ga=87.0$). The angle of the zigzag plane is indicated at the bottom row. The wall, not displayed for the sake of clarity, is placed along plane $yz$ at $y$=-4, -2 and -1 for $L=$4, 2 and 1, respectively.}\label{fig:3D_xy_short_wall}
\end{figure}

In order to compare the different cases together, Fig.~\ref{fig:xz_short_wall} shows the vertical distance from the bubble centroid to the wall, $z-z_w$ as a function of $x$ coordinate. In this way, $z-z_w$=0 indicates the wall edge, while in the experiments with a long wall, $z_w$= 0. The results for the rectilinear regime are presented in Fig.~\ref{fig:xz_short_wall}(a). Notice that, when the short wall is used (dashed lines), the migration effect starts before reaching the wall, at $z<z_w$, and the bubble displacement is much larger than that observed with a long wall ($z_w$= 0). This effect is more pronounced as $L$ decreases. The same comparison is made for the zigzag and the spiral regimes in Figs. ~\ref{fig:xz_short_wall}(b) and (c), respectively. In these cases, averaged paths are plotted for the sake of clarity. The same qualitative picture as the one described for the rectilinear regime can be observed, but notice that the migration occurs at even lower vertical positions. In the zigzag case, depending on the phase of the zigzagging movement when the bubble reaches the wall edge, the induced migration effect can take place even sooner. The early migration is more notable in the spiral regime due to the three-dimensional nature of the paths, being the bubbles bigger and more deformed than in the other regimes. It is worth mentioning that both unstable cases with the short wall are not quite reproducible since the final migration effect depends on the exact relative position between the bubble and the wall, which is different for each experiment.

Since Figs.~\ref{fig:xz_short_wall}(b) and (c) show the average bubble paths of the unstable regimes, it is interesting to explore the complete rising trajectories within these unstable regimes when the short wall is used. In this regard, Figs.~\ref{fig:3D_xy_short_wall}(a)-(c) show the three-dimensional rising path corresponding to the planar zigzagging regime. Before reaching the wall, the bubble freely rises in a zigzag regime. Notice that, just before finding the wall edge, the lateral displacement from the wall takes place while keeping the zigzag motion. As in its long-wall counterpart (see Fig.\ref{fig:trajs_zigzag}), this effect is particularly clear for the smallest value of $L$ (Fig.~\ref{fig:3D_xy_short_wall}c). Moreover, as shown in the corresponding top views displayed in Figs.~\ref{fig:3D_xy_short_wall}(d)-(f), the established zigzagging plane was completely random, even for $L$=1, differently from the long wall experiments, in which the wall sets a boundary condition imposing a perpendicular zigzagging plane when the bubble is close enough. Except in the local region close to the wall edge, the wall has been found not to modify the oscillation frequency or the oscillating amplitude. 
\begin{figure}[t!]
    \centering
    \includegraphics[width=1\textwidth]{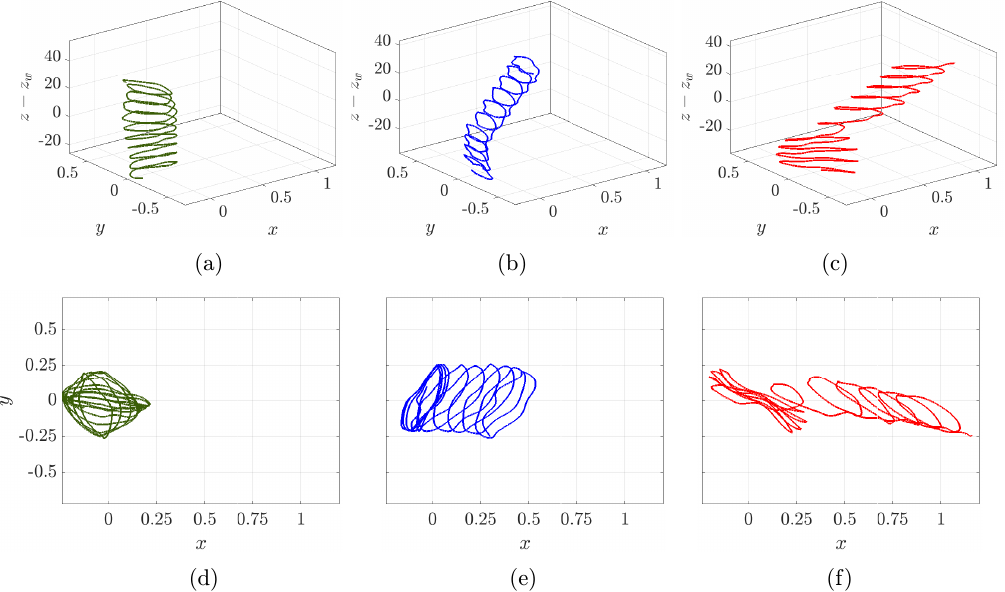}
    \caption{3D (top row) and xy (bottom row) bubble paths for spiral regime using a short wall and (a), (d) $L$=4 (green, $Bo=12.34,~Ga=128.1$), (b), (e) $L$=2 (blue, $Bo=12.33,~Ga=128.5$), and (c), (f) $L$=1 (red, $Bo=12.13,~Ga=124.7$). The wall, not displayed for the sake of clarity, is placed along plane $yz$ at $y$=-4, -2 and -1 for $L=$4, 2 and 1, respectively.}\label{fig:spiral_short_wall}
\end{figure}

With regard to the spiraling regime, Fig.\ref{fig:spiral_short_wall} shows the three-dimensional and top views of the bubble path when a short wall is placed into the tank. The bubble clearly migrates from the wall just before reaching its edge. Compared to its long-wall counterpart (Fig.\ref{fig:trajs_spiral}) notice that, when the bubble is released, it freely rises without any constraint, differently from the long-wall case. Thus, paths are more spiral-like than in the case with the long wall. This fact can be clearly observed in the top views (Figs.\ref{fig:spiral_short_wall}d-f), where the path horizontal projections show a more elliptical shape than in the cases with the long wall, which tend to evolve into flattened spirals. While in the long-wall case the wall imposes a restriction for the spiral to develop, in the current case the final regime is completely established before reaching the wall.

\section{Conclusions}\label{sec:conclusions}
In this work, the rising dynamics of bubbles with high $Bo$ in the presence of a rigid solid wall are experimentally analyzed. Three different rising regimes, namely rectilinear, planar zigzagging, and spiraling are explored. The initial horizontal distance between the wall and the bubble centroid is also varied. Some discrepancies have been found with previous works studying spherical bubbles, which justifies our focus on the high $Bo$ regime, typical of ellipsoidal bubbles, whose deformation leads to complex asymmetric vortex structures in the flow, enhancing the wall effect. 

The main observed effect for all cases and bubble-wall initial distances is a transverse migration of the bubble away from the wall that increases with bubble-wall initial proximity. In the rectilinear and zigzagging cases, the bubble keeps a similar behavior to the unbounded cases, i.e. they maintain the expected trajectories, that are only altered by this transverse motion. When comparing the impact of the wall on the average paths of the different regimes, it can be observed that the largest migration effect takes place for the zigzagging regime, being the lowest for the stable case. Moreover, in the spiraling regime, besides the bubble migration from the wall, a change in the rising path is promoted when the wall is initially close to the bubble. In particular, the limitation imposed by the wall prevents the bubble from following a complete spiral path from the very beginning, promoting a more zigzag-like trajectory that eventually leads to a flattened spiraling motion.

Unstable regimes seem to be established earlier due to the wall presence, which means that the wall triggers instability. A slight increase in amplitude due to this destabilizing effect was also observed in the unstable regimes, as found numerically by \citet
{Zhang2020}.
However, when the wall is sufficiently close to the bubble, the zigzagging plane is established normal to its surface. This fact indicates that the wall can impose an initial condition on the bubble trajectory development as soon as the bubble is generated.

Regarding the vertical bubble velocity, the presence of the wall promotes a larger initial acceleration which shortens the time necessary for the bubble to reach its terminal velocity. Besides, this velocity decreases with the proximity of the wall due to an increase in the drag force. Nevertheless, this velocity drop is only observable in our experiments when the bubble and wall are very close or when the spiraling regime is established. Additionally, oscillations in the vertical velocity have been obtained in the zigzagging regime, as reported numerically by \citet{Cano-Lozano2016}. Moreover, the experimental Reynolds number associated to the terminal velocity and corresponding to the unbounded case are in good agreement with the ones obtained from correlations valid for clean bubbles. This indicates that contamination, if present, is not relevant in our experiments. 

In addition, it has been observed that, when the wall edge is placed far above the bubble injection point, the bubble displacement starts before it reaches the wall, and deviates longer distances from the surface. Focusing on the planar zigzagging case, since the oscillating plane is already established before the bubble reaches the wall position, this plane is not altered, only the migration effect is added to the already developed trajectory. Indeed, the zigzagging plane is established arbitrarily, as happens in the long-wall experiments when the initial wall distance to the bubble is sufficiently large. For the spiraling case, a later wall interaction allows for more spiral-like paths as the wall spatial limitation near to the injector is eliminated.

\begin{acknowledgments}
This work was supported by the coordinated project, PID2020-115961RB-C31, PID2020-115961RB-C32 and PID2020-115961RA-C33, financed by MCIN/AEI/10.13039/501100011033. C. Estepa-Cantero would like to thank the Spanish Ministry of Universities for the financial support provided by the Fellowship FPU20/02197.
\end{acknowledgments}


\section*{References}
\begin{thebibliography}{65}%
\makeatletter
\providecommand \@ifxundefined [1]{%
 \@ifx{#1\undefined}
}%
\providecommand \@ifnum [1]{%
 \ifnum #1\expandafter \@firstoftwo
 \else \expandafter \@secondoftwo
 \fi
}%
\providecommand \@ifx [1]{%
 \ifx #1\expandafter \@firstoftwo
 \else \expandafter \@secondoftwo
 \fi
}%
\providecommand \natexlab [1]{#1}%
\providecommand \enquote  [1]{``#1''}%
\providecommand \bibnamefont  [1]{#1}%
\providecommand \bibfnamefont [1]{#1}%
\providecommand \citenamefont [1]{#1}%
\providecommand \href@noop [0]{\@secondoftwo}%
\providecommand \href [0]{\begingroup \@sanitize@url \@href}%
\providecommand \@href[1]{\@@startlink{#1}\@@href}%
\providecommand \@@href[1]{\endgroup#1\@@endlink}%
\providecommand \@sanitize@url [0]{\catcode `\\12\catcode `\$12\catcode
  `\&12\catcode `\#12\catcode `\^12\catcode `\_12\catcode `\%12\relax}%
\providecommand \@@startlink[1]{}%
\providecommand \@@endlink[0]{}%
\providecommand \url  [0]{\begingroup\@sanitize@url \@url }%
\providecommand \@url [1]{\endgroup\@href {#1}{\urlprefix }}%
\providecommand \urlprefix  [0]{URL }%
\providecommand \Eprint [0]{\href }%
\providecommand \doibase [0]{http://dx.doi.org/}%
\providecommand \selectlanguage [0]{\@gobble}%
\providecommand \bibinfo  [0]{\@secondoftwo}%
\providecommand \bibfield  [0]{\@secondoftwo}%
\providecommand \translation [1]{[#1]}%
\providecommand \BibitemOpen [0]{}%
\providecommand \bibitemStop [0]{}%
\providecommand \bibitemNoStop [0]{.\EOS\space}%
\providecommand \EOS [0]{\spacefactor3000\relax}%
\providecommand \BibitemShut  [1]{\csname bibitem#1\endcsname}%
\let\auto@bib@innerbib\@empty
\bibitem [{\citenamefont {Dean}\ \emph {et~al.}(2018)\citenamefont {Dean},
  \citenamefont {Middelburg}, \citenamefont {R{\"o}ckmann}, \citenamefont
  {Aerts}, \citenamefont {Blauw}, \citenamefont {Egger}, \citenamefont
  {Jetten}, \citenamefont {de~Jong}, \citenamefont {Meisel}, \citenamefont
  {Rasigraf} \emph {et~al.}}]{dean2018methane}%
  \BibitemOpen
  \bibfield  {author} {\bibinfo {author} {\bibfnamefont {J.~F.}\ \bibnamefont
  {Dean}}, \bibinfo {author} {\bibfnamefont {J.~J.}\ \bibnamefont
  {Middelburg}}, \bibinfo {author} {\bibfnamefont {T.}~\bibnamefont
  {R{\"o}ckmann}}, \bibinfo {author} {\bibfnamefont {R.}~\bibnamefont {Aerts}},
  \bibinfo {author} {\bibfnamefont {L.~G.}\ \bibnamefont {Blauw}}, \bibinfo
  {author} {\bibfnamefont {M.}~\bibnamefont {Egger}}, \bibinfo {author}
  {\bibfnamefont {M.~S.}\ \bibnamefont {Jetten}}, \bibinfo {author}
  {\bibfnamefont {A.~E.}\ \bibnamefont {de~Jong}}, \bibinfo {author}
  {\bibfnamefont {O.~H.}\ \bibnamefont {Meisel}}, \bibinfo {author}
  {\bibfnamefont {O.}~\bibnamefont {Rasigraf}},  \emph {et~al.},\ }\bibfield
  {title} {\enquote {\bibinfo {title} {Methane feedbacks to the global climate
  system in a warmer world},}\ }\href@noop {} {\bibfield  {journal} {\bibinfo
  {journal} {Rev. Geophys.}\ }\textbf {\bibinfo {volume} {56}},\ \bibinfo
  {pages} {207--250} (\bibinfo {year} {2018})}\BibitemShut {NoStop}%
\bibitem [{\citenamefont {Liu}\ \emph {et~al.}(2015)\citenamefont {Liu},
  \citenamefont {Cashman}, \citenamefont {Rust},\ and\ \citenamefont
  {Gislason}}]{liu2015role}%
  \BibitemOpen
  \bibfield  {author} {\bibinfo {author} {\bibfnamefont {E.}~\bibnamefont
  {Liu}}, \bibinfo {author} {\bibfnamefont {K.}~\bibnamefont {Cashman}},
  \bibinfo {author} {\bibfnamefont {A.}~\bibnamefont {Rust}}, \ and\ \bibinfo
  {author} {\bibfnamefont {S.}~\bibnamefont {Gislason}},\ }\bibfield  {title}
  {\enquote {\bibinfo {title} {The role of bubbles in generating fine ash
  during hydromagmatic eruptions},}\ }\href@noop {} {\bibfield  {journal}
  {\bibinfo  {journal} {Geology}\ }\textbf {\bibinfo {volume} {43}},\ \bibinfo
  {pages} {239--242} (\bibinfo {year} {2015})}\BibitemShut {NoStop}%
\bibitem [{\citenamefont {Rodríguez-Rodríguez}\ \emph
  {et~al.}(2015)\citenamefont {Rodríguez-Rodríguez}, \citenamefont {Sevilla},
  \citenamefont {Martínez-Bazán},\ and\ \citenamefont
  {Gordillo}}]{Rodriguez-Rodriguez2015}%
  \BibitemOpen
  \bibfield  {author} {\bibinfo {author} {\bibfnamefont {J.}~\bibnamefont
  {Rodríguez-Rodríguez}}, \bibinfo {author} {\bibfnamefont {A.}~\bibnamefont
  {Sevilla}}, \bibinfo {author} {\bibfnamefont {C.}~\bibnamefont
  {Martínez-Bazán}}, \ and\ \bibinfo {author} {\bibfnamefont {J.~M.}\
  \bibnamefont {Gordillo}},\ }\bibfield  {title} {\enquote {\bibinfo {title}
  {Generation of microbubbles with applications to industry and medicine},}\
  }\href@noop {} {\bibfield  {journal} {\bibinfo  {journal} {Annual Review of
  Fluid Mechanics}\ }\textbf {\bibinfo {volume} {47}},\ \bibinfo {pages}
  {405--429} (\bibinfo {year} {2015})}\BibitemShut {NoStop}%
\bibitem [{\citenamefont {Rajapakse}\ \emph {et~al.}(2022)\citenamefont
  {Rajapakse}, \citenamefont {Zargar}, \citenamefont {Sen},\ and\ \citenamefont
  {Khiadani}}]{rajapakse2022effects}%
  \BibitemOpen
  \bibfield  {author} {\bibinfo {author} {\bibfnamefont {N.}~\bibnamefont
  {Rajapakse}}, \bibinfo {author} {\bibfnamefont {M.}~\bibnamefont {Zargar}},
  \bibinfo {author} {\bibfnamefont {T.}~\bibnamefont {Sen}}, \ and\ \bibinfo
  {author} {\bibfnamefont {M.}~\bibnamefont {Khiadani}},\ }\bibfield  {title}
  {\enquote {\bibinfo {title} {Effects of influent physicochemical
  characteristics on air dissolution, bubble size and rise velocity in
  dissolved air flotation: A review},}\ }\href@noop {} {\bibfield  {journal}
  {\bibinfo  {journal} {Sep. Purif. Technol.}\ }\textbf {\bibinfo {volume}
  {289}},\ \bibinfo {pages} {120772} (\bibinfo {year} {2022})}\BibitemShut
  {NoStop}%
\bibitem [{\citenamefont {Shaw}\ and\ \citenamefont {Deike}(2022)}]{Shaw2022}%
  \BibitemOpen
  \bibfield  {author} {\bibinfo {author} {\bibfnamefont {D.}~\bibnamefont
  {Shaw}}\ and\ \bibinfo {author} {\bibfnamefont {L.}~\bibnamefont {Deike}},\
  }\bibfield  {title} {\enquote {\bibinfo {title} {Microplastic transport by
  bursting bubbles},}\ }\href@noop {} {\bibfield  {journal} {\bibinfo
  {journal} {Bulletin of the American Physical Society}\ } (\bibinfo {year}
  {2022})}\BibitemShut {NoStop}%
\bibitem [{\citenamefont {Wang}, \citenamefont {Wang},\ and\ \citenamefont
  {Liu}(2022)}]{Wang2022}%
  \BibitemOpen
  \bibfield  {author} {\bibinfo {author} {\bibfnamefont {H.}~\bibnamefont
  {Wang}}, \bibinfo {author} {\bibfnamefont {K.}~\bibnamefont {Wang}}, \ and\
  \bibinfo {author} {\bibfnamefont {G.}~\bibnamefont {Liu}},\ }\bibfield
  {title} {\enquote {\bibinfo {title} {Drag reduction by gas lubrication with
  bubbles},}\ }\href@noop {} {\bibfield  {journal} {\bibinfo  {journal} {Ocean
  Eng.}\ }\textbf {\bibinfo {volume} {258}},\ \bibinfo {pages} {111833}
  (\bibinfo {year} {2022})}\BibitemShut {NoStop}%
\bibitem [{\citenamefont {Zhao}, \citenamefont {Wright},\ and\ \citenamefont
  {Goertz}(2023)}]{Zhao2023}%
  \BibitemOpen
  \bibfield  {author} {\bibinfo {author} {\bibfnamefont {X.}~\bibnamefont
  {Zhao}}, \bibinfo {author} {\bibfnamefont {A.}~\bibnamefont {Wright}}, \ and\
  \bibinfo {author} {\bibfnamefont {D.~E.}\ \bibnamefont {Goertz}},\ }\bibfield
   {title} {\enquote {\bibinfo {title} {An optical and acoustic investigation
  of microbubble cavitation in small channels under therapeutic ultrasound
  conditions},}\ }\href@noop {} {\bibfield  {journal} {\bibinfo  {journal}
  {Ultrason. Sonochem.}\ }\textbf {\bibinfo {volume} {93}},\ \bibinfo {pages}
  {106291} (\bibinfo {year} {2023})}\BibitemShut {NoStop}%
\bibitem [{\citenamefont {Lehmann}, \citenamefont {H{\"a}usl},\ and\
  \citenamefont {Gekle}(2023)}]{Lehmann2023}%
  \BibitemOpen
  \bibfield  {author} {\bibinfo {author} {\bibfnamefont {M.}~\bibnamefont
  {Lehmann}}, \bibinfo {author} {\bibfnamefont {F.~P.}\ \bibnamefont
  {H{\"a}usl}}, \ and\ \bibinfo {author} {\bibfnamefont {S.}~\bibnamefont
  {Gekle}},\ }\bibfield  {title} {\enquote {\bibinfo {title} {Modeling of
  vertical microplastic transport by rising bubbles},}\ }\href@noop {}
  {\bibfield  {journal} {\bibinfo  {journal} {Microplastics Nanoplastics}\
  }\textbf {\bibinfo {volume} {3:4}},\ \bibinfo {pages} {1--6} (\bibinfo {year}
  {2023})}\BibitemShut {NoStop}%
\bibitem [{\citenamefont {Gao}\ \emph {et~al.}(2023)\citenamefont {Gao},
  \citenamefont {Yang}, \citenamefont {Qi}, \citenamefont {Guo},\ and\
  \citenamefont {Zhang}}]{Gao2023}%
  \BibitemOpen
  \bibfield  {author} {\bibinfo {author} {\bibfnamefont {D.}~\bibnamefont
  {Gao}}, \bibinfo {author} {\bibfnamefont {J.}~\bibnamefont {Yang}}, \bibinfo
  {author} {\bibfnamefont {Y.}~\bibnamefont {Qi}}, \bibinfo {author}
  {\bibfnamefont {C.}~\bibnamefont {Guo}}, \ and\ \bibinfo {author}
  {\bibfnamefont {H.}~\bibnamefont {Zhang}},\ }\bibfield  {title} {\enquote
  {\bibinfo {title} {Review and perspectives on {CO}2 bubble dynamic
  characteristics in different liquids during carbon capture, utilization, and
  storage process},}\ }\href@noop {} {\bibfield  {journal} {\bibinfo  {journal}
  {Energy Fuels}\ }\textbf {\bibinfo {volume} {37}},\ \bibinfo {pages} {58--73}
  (\bibinfo {year} {2023})}\BibitemShut {NoStop}%
\bibitem [{\citenamefont {Ri}\ \emph {et~al.}(2023)\citenamefont {Ri},
  \citenamefont {Pang}, \citenamefont {Bai}, \citenamefont {Xu}, \citenamefont
  {Xu}, \citenamefont {Ri}, \citenamefont {Yao},\ and\ \citenamefont
  {Greenwald}}]{Ri2023}%
  \BibitemOpen
  \bibfield  {author} {\bibinfo {author} {\bibfnamefont {J.}~\bibnamefont
  {Ri}}, \bibinfo {author} {\bibfnamefont {N.}~\bibnamefont {Pang}}, \bibinfo
  {author} {\bibfnamefont {S.}~\bibnamefont {Bai}}, \bibinfo {author}
  {\bibfnamefont {J.}~\bibnamefont {Xu}}, \bibinfo {author} {\bibfnamefont
  {L.}~\bibnamefont {Xu}}, \bibinfo {author} {\bibfnamefont {S.}~\bibnamefont
  {Ri}}, \bibinfo {author} {\bibfnamefont {Y.}~\bibnamefont {Yao}}, \ and\
  \bibinfo {author} {\bibfnamefont {S.~E.}\ \bibnamefont {Greenwald}},\
  }\bibfield  {title} {\enquote {\bibinfo {title} {{Three-dimensional numerical
  analysis of wall stress induced by asymmetric oscillation of microbubble
  trains inside micro-vessels}},}\ }\href {\doibase 10.1063/5.0134922}
  {\bibfield  {journal} {\bibinfo  {journal} {Phys.Fluids}\ }\textbf
  {\bibinfo {volume} {35}},\ \bibinfo {pages} {011904} (\bibinfo {year}
  {2023})}\BibitemShut {NoStop}%
\bibitem [{\citenamefont {Clift}, \citenamefont {Grace},\ and\ \citenamefont
  {Weber}(1978)}]{Clift1978}%
  \BibitemOpen
  \bibfield  {author} {\bibinfo {author} {\bibfnamefont {R.}~\bibnamefont
  {Clift}}, \bibinfo {author} {\bibfnamefont {J.~R.}\ \bibnamefont {Grace}}, \
  and\ \bibinfo {author} {\bibfnamefont {M.~E.}\ \bibnamefont {Weber}},\
  }\href@noop {} {\emph {\bibinfo {title} {Bubbles, Drops and Particles}}}\
  (\bibinfo  {publisher} {Academic Press, INC.},\ \bibinfo {year}
  {1978})\BibitemShut {NoStop}%
\bibitem [{\citenamefont {Marusic}\ and\ \citenamefont
  {Broomhall}(2021)}]{marusic2021leonardo}%
  \BibitemOpen
  \bibfield  {author} {\bibinfo {author} {\bibfnamefont {I.}~\bibnamefont
  {Marusic}}\ and\ \bibinfo {author} {\bibfnamefont {S.}~\bibnamefont
  {Broomhall}},\ }\bibfield  {title} {\enquote {\bibinfo {title} {Leonardo da
  vinci and fluid mechanics},}\ }\href@noop {} {\bibfield  {journal} {\bibinfo
  {journal} {Annu. Rev. Fluid Mech.}\ }\textbf {\bibinfo {volume} {53}},\
  \bibinfo {pages} {1--25} (\bibinfo {year} {2021})}\BibitemShut {NoStop}%
\bibitem [{\citenamefont {Yang}, \citenamefont {Prosperetti},\ and\
  \citenamefont {Takagi}(2003)}]{Yang2003}%
  \BibitemOpen
  \bibfield  {author} {\bibinfo {author} {\bibfnamefont {B.}~\bibnamefont
  {Yang}}, \bibinfo {author} {\bibfnamefont {A.}~\bibnamefont {Prosperetti}}, \
  and\ \bibinfo {author} {\bibfnamefont {S.}~\bibnamefont {Takagi}},\
  }\bibfield  {title} {\enquote {\bibinfo {title} {The transient rise of a
  bubble subject to shape or volume changes},}\ }\href@noop {} {\bibfield
  {journal} {\bibinfo  {journal} {Phys.Fluids}\ }\textbf {\bibinfo {volume}
  {15}},\ \bibinfo {pages} {2640--2648} (\bibinfo {year} {2003})}\BibitemShut
  {NoStop}%
\bibitem [{\citenamefont {Blanco}\ and\ \citenamefont
  {Magnaudet}(1995)}]{blanco1995structure}%
  \BibitemOpen
  \bibfield  {author} {\bibinfo {author} {\bibfnamefont {A.}~\bibnamefont
  {Blanco}}\ and\ \bibinfo {author} {\bibfnamefont {J.}~\bibnamefont
  {Magnaudet}},\ }\bibfield  {title} {\enquote {\bibinfo {title} {The structure
  of the axisymmetric high-reynolds number flow around an ellipsoidal bubble of
  fixed shape},}\ }\href@noop {} {\bibfield  {journal} {\bibinfo  {journal}
  {Phys. Fluids}\ }\textbf {\bibinfo {volume} {7}},\ \bibinfo {pages}
  {1265--1274} (\bibinfo {year} {1995})}\BibitemShut {NoStop}%
\bibitem [{\citenamefont {Magnaudet}\ and\ \citenamefont
  {Eames}(2000)}]{Magnaudet2000}%
  \BibitemOpen
  \bibfield  {author} {\bibinfo {author} {\bibfnamefont {J.}~\bibnamefont
  {Magnaudet}}\ and\ \bibinfo {author} {\bibfnamefont {I.}~\bibnamefont
  {Eames}},\ }\bibfield  {title} {\enquote {\bibinfo {title} {The motion of
  high-reynolds-number bubbles in inhomogeneus flows},}\ }\href@noop {}
  {\bibfield  {journal} {\bibinfo  {journal} {Annu. Rev. Fluid Mech.}\ }\textbf
  {\bibinfo {volume} {32}},\ \bibinfo {pages} {659--708} (\bibinfo {year}
  {2000})}\BibitemShut {NoStop}%
\bibitem [{\citenamefont {Mougin}(2002)}]{mougin2002paht}%
  \BibitemOpen
  \bibfield  {author} {\bibinfo {author} {\bibfnamefont {G.}~\bibnamefont
  {Mougin}},\ }\bibfield  {title} {\enquote {\bibinfo {title} {Paht instability
  of a rising bubble},}\ }\href@noop {} {\bibfield  {journal} {\bibinfo
  {journal} {Phys. Rev. Lett.}\ }\textbf {\bibinfo {volume} {88}},\ \bibinfo
  {pages} {041502} (\bibinfo {year} {2002})}\BibitemShut {NoStop}%
\bibitem [{\citenamefont {Mougin}\ and\ \citenamefont
  {Magnaudet}(2006)}]{mougin2006wake}%
  \BibitemOpen
  \bibfield  {author} {\bibinfo {author} {\bibfnamefont {G.}~\bibnamefont
  {Mougin}}\ and\ \bibinfo {author} {\bibfnamefont {J.}~\bibnamefont
  {Magnaudet}},\ }\bibfield  {title} {\enquote {\bibinfo {title} {Wake-induced
  forces and torques on a zigzagging/spiralling bubble},}\ }\href@noop {}
  {\bibfield  {journal} {\bibinfo  {journal} {J. Fluid Mech.}\ }\textbf
  {\bibinfo {volume} {567}},\ \bibinfo {pages} {185--194} (\bibinfo {year}
  {2006})}\BibitemShut {NoStop}%
\bibitem [{\citenamefont {Magnaudet}\ and\ \citenamefont
  {Mougin}(2007)}]{magnaudet2007wake}%
  \BibitemOpen
  \bibfield  {author} {\bibinfo {author} {\bibfnamefont {J.}~\bibnamefont
  {Magnaudet}}\ and\ \bibinfo {author} {\bibfnamefont {G.}~\bibnamefont
  {Mougin}},\ }\bibfield  {title} {\enquote {\bibinfo {title} {Wake instability
  of a fixed spheroidal bubble},}\ }\href@noop {} {\bibfield  {journal}
  {\bibinfo  {journal} {J. Fluid Mech.}\ }\textbf {\bibinfo {volume} {572}},\
  \bibinfo {pages} {311--337} (\bibinfo {year} {2007})}\BibitemShut {NoStop}%
\bibitem [{\citenamefont {Tripathi}, \citenamefont {Sahu},\ and\ \citenamefont
  {Govindarajan}(2015)}]{tripathi2015dynamics}%
  \BibitemOpen
  \bibfield  {author} {\bibinfo {author} {\bibfnamefont {M.~K.}\ \bibnamefont
  {Tripathi}}, \bibinfo {author} {\bibfnamefont {K.~C.}\ \bibnamefont {Sahu}},
  \ and\ \bibinfo {author} {\bibfnamefont {R.}~\bibnamefont {Govindarajan}},\
  }\bibfield  {title} {\enquote {\bibinfo {title} {Dynamics of an initially
  spherical bubble rising in quiescent liquid},}\ }\href@noop {} {\bibfield
  {journal} {\bibinfo  {journal} {Nat. {C}ommun.}\ }\textbf {\bibinfo {volume}
  {6}},\ \bibinfo {pages} {6268} (\bibinfo {year} {2015})}\BibitemShut
  {NoStop}%
\bibitem [{\citenamefont {Cano-Lozano}\ \emph {et~al.}(2015)\citenamefont
  {Cano-Lozano}, \citenamefont {Bolaños-Jiménez}, \citenamefont
  {Gutiérrez-Montes},\ and\ \citenamefont
  {Martínez-Bazán}}]{Cano-Lozano2015}%
  \BibitemOpen
  \bibfield  {author} {\bibinfo {author} {\bibfnamefont {J.~C.}\ \bibnamefont
  {Cano-Lozano}}, \bibinfo {author} {\bibfnamefont {R.}~\bibnamefont
  {Bolaños-Jiménez}}, \bibinfo {author} {\bibfnamefont {C.}~\bibnamefont
  {Gutiérrez-Montes}}, \ and\ \bibinfo {author} {\bibfnamefont
  {C.}~\bibnamefont {Martínez-Bazán}},\ }\bibfield  {title} {\enquote
  {\bibinfo {title} {The use of volume of fluid technique to analyze multiphase
  flows: Specific case of bubble rising in still liquids},}\ }\href@noop {}
  {\bibfield  {journal} {\bibinfo  {journal} {Appl. Math. Model.}\ }\textbf
  {\bibinfo {volume} {39}},\ \bibinfo {pages} {3290--3305} (\bibinfo {year}
  {2015})}\BibitemShut {NoStop}%
\bibitem [{\citenamefont {Cano-Lozano}\ \emph
  {et~al.}(2016{\natexlab{a}})\citenamefont {Cano-Lozano}, \citenamefont
  {Martínez-Bazán}, \citenamefont {Magnaudet},\ and\ \citenamefont
  {Tchoufag}}]{Cano-Lozano2016}%
  \BibitemOpen
  \bibfield  {author} {\bibinfo {author} {\bibfnamefont {J.~C.}\ \bibnamefont
  {Cano-Lozano}}, \bibinfo {author} {\bibfnamefont {C.}~\bibnamefont
  {Martínez-Bazán}}, \bibinfo {author} {\bibfnamefont {J.}~\bibnamefont
  {Magnaudet}}, \ and\ \bibinfo {author} {\bibfnamefont {J.}~\bibnamefont
  {Tchoufag}},\ }\bibfield  {title} {\enquote {\bibinfo {title} {Paths and
  wakes of deformable nearly spheroidal rising bubbles close to the transition
  to path instability},}\ }\href@noop {} {\bibfield  {journal} {\bibinfo
  {journal} {Phys. Rev. Fluids}\ }\textbf {\bibinfo {volume} {1}},\ \bibinfo
  {pages} {1--30} (\bibinfo {year} {2016}{\natexlab{a}})}\BibitemShut {NoStop}%
\bibitem [{\citenamefont {Zhang}\ \emph {et~al.}(2021)\citenamefont {Zhang},
  \citenamefont {Peng}, \citenamefont {Shao},\ and\ \citenamefont
  {Deng}}]{Zhang2021}%
  \BibitemOpen
  \bibfield  {author} {\bibinfo {author} {\bibfnamefont {L.}~\bibnamefont
  {Zhang}}, \bibinfo {author} {\bibfnamefont {K.}~\bibnamefont {Peng}},
  \bibinfo {author} {\bibfnamefont {X.}~\bibnamefont {Shao}}, \ and\ \bibinfo
  {author} {\bibfnamefont {J.}~\bibnamefont {Deng}},\ }\bibfield  {title}
  {\enquote {\bibinfo {title} {Direct numerical simulation of deformable rising
  bubbles at low reynolds numbers},}\ }\href {\doibase 10.1063/5.0072725}
  {\bibfield  {journal} {\bibinfo  {journal} {Phys.Fluids}\ }\textbf
  {\bibinfo {volume} {33}},\ \bibinfo {pages} {113309} (\bibinfo {year}
  {2021})}\BibitemShut {NoStop}%
\bibitem [{\citenamefont {Bonnefis}, \citenamefont {Fabre},\ and\ \citenamefont
  {Magnaudet}(2023)}]{bonnefis2023and}%
  \BibitemOpen
  \bibfield  {author} {\bibinfo {author} {\bibfnamefont {P.}~\bibnamefont
  {Bonnefis}}, \bibinfo {author} {\bibfnamefont {D.}~\bibnamefont {Fabre}}, \
  and\ \bibinfo {author} {\bibfnamefont {J.}~\bibnamefont {Magnaudet}},\
  }\bibfield  {title} {\enquote {\bibinfo {title} {When, how, and why the path
  of an air bubble rising in pure water becomes unstable},}\ }\href@noop {}
  {\bibfield  {journal} {\bibinfo  {journal} {Proceedings of the National
  Academy of Sciences}\ }\textbf {\bibinfo {volume} {120}},\ \bibinfo {pages}
  {e2300897120} (\bibinfo {year} {2023})}\BibitemShut {NoStop}%
\bibitem [{\citenamefont {Haberman}\ and\ \citenamefont
  {Morton}(1953)}]{haberman1953experimental}%
  \BibitemOpen
  \bibfield  {author} {\bibinfo {author} {\bibfnamefont {W.~L.}\ \bibnamefont
  {Haberman}}\ and\ \bibinfo {author} {\bibfnamefont {R.}~\bibnamefont
  {Morton}},\ }\href@noop {} {\emph {\bibinfo {title} {An experimental
  investigation of the drag and shape of air bubbles rising in various
  liquids}}}\ (\bibinfo  {publisher} {David W. Taylor Model Basin Washington,
  DC},\ \bibinfo {year} {1953})\BibitemShut {NoStop}%
\bibitem [{\citenamefont {Bhaga}\ and\ \citenamefont
  {Weber}(1981)}]{Bhaga1981}%
  \BibitemOpen
  \bibfield  {author} {\bibinfo {author} {\bibfnamefont {D.}~\bibnamefont
  {Bhaga}}\ and\ \bibinfo {author} {\bibfnamefont {M.~E.}\ \bibnamefont
  {Weber}},\ }\bibfield  {title} {\enquote {\bibinfo {title} {Bubbles in
  viscous liquids: shapes, wakes and velocities},}\ }\href@noop {} {\bibfield
  {journal} {\bibinfo  {journal} {J. Fluid Mech.}\ }\textbf {\bibinfo {volume}
  {105}},\ \bibinfo {pages} {61--85} (\bibinfo {year} {1981})}\BibitemShut
  {NoStop}%
\bibitem [{\citenamefont {Duineveld}(1995)}]{Duineveld1995}%
  \BibitemOpen
  \bibfield  {author} {\bibinfo {author} {\bibfnamefont {P.~C.}\ \bibnamefont
  {Duineveld}},\ }\bibfield  {title} {\enquote {\bibinfo {title} {The rise
  velocity and shape of bubbles in pure water at high reynolds number},}\
  }\href@noop {} {\bibfield  {journal} {\bibinfo  {journal} {J. Fluid Mech.}\
  }\textbf {\bibinfo {volume} {292}},\ \bibinfo {pages} {325–332} (\bibinfo
  {year} {1995})}\BibitemShut {NoStop}%
\bibitem [{\citenamefont {Maxworthy}\ \emph {et~al.}(1996)\citenamefont
  {Maxworthy}, \citenamefont {Gnann}, \citenamefont {Kürten},\ and\
  \citenamefont {Durst}}]{Maxworthy1996}%
  \BibitemOpen
  \bibfield  {author} {\bibinfo {author} {\bibfnamefont {T.}~\bibnamefont
  {Maxworthy}}, \bibinfo {author} {\bibfnamefont {C.}~\bibnamefont {Gnann}},
  \bibinfo {author} {\bibfnamefont {M.}~\bibnamefont {Kürten}}, \ and\
  \bibinfo {author} {\bibfnamefont {F.}~\bibnamefont {Durst}},\ }\bibfield
  {title} {\enquote {\bibinfo {title} {Experiments on the rise of air bubbles
  in clean viscous liquids},}\ }\href@noop {} {\bibfield  {journal} {\bibinfo
  {journal} {Journal of Fluid Mechanics}\ }\textbf {\bibinfo {volume} {321}},\
  \bibinfo {pages} {421--441} (\bibinfo {year} {1996})}\BibitemShut {NoStop}%
\bibitem [{\citenamefont {Ellingsen}\ and\ \citenamefont
  {Risso}(2001)}]{Ellingsen2001}%
  \BibitemOpen
  \bibfield  {author} {\bibinfo {author} {\bibfnamefont {K.}~\bibnamefont
  {Ellingsen}}\ and\ \bibinfo {author} {\bibfnamefont {F.}~\bibnamefont
  {Risso}},\ }\bibfield  {title} {\enquote {\bibinfo {title} {On the rise of an
  ellipsoidal bubble in water: Oscillatory paths and liquid-induced
  velocity},}\ }\href@noop {} {\bibfield  {journal} {\bibinfo  {journal} {J.
  Fluid Mech.}\ }\textbf {\bibinfo {volume} {440}},\ \bibinfo {pages}
  {235--268} (\bibinfo {year} {2001})}\BibitemShut {NoStop}%
\bibitem [{\citenamefont {Vries}, \citenamefont {Biesheuvel},\ and\
  \citenamefont {Wijngaarden}(2003)}]{DeVries2003}%
  \BibitemOpen
  \bibfield  {author} {\bibinfo {author} {\bibfnamefont {A.~W.~D.}\
  \bibnamefont {Vries}}, \bibinfo {author} {\bibfnamefont {A.}~\bibnamefont
  {Biesheuvel}}, \ and\ \bibinfo {author} {\bibfnamefont {L.~V.}\ \bibnamefont
  {Wijngaarden}},\ }\bibfield  {title} {\enquote {\bibinfo {title} {Notes on
  the path and wake of a gas bubble rising in pure water},}\ }\href@noop {}
  {\bibfield  {journal} {\bibinfo  {journal} {Int. J. Multiph. Flow}\ }\textbf
  {\bibinfo {volume} {28}},\ \bibinfo {pages} {1823--1835} (\bibinfo {year}
  {2003})}\BibitemShut {NoStop}%
\bibitem [{\citenamefont {Shew}, \citenamefont {Poncet},\ and\ \citenamefont
  {Pinton}(2006)}]{Shew2006}%
  \BibitemOpen
  \bibfield  {author} {\bibinfo {author} {\bibfnamefont {W.~L.}\ \bibnamefont
  {Shew}}, \bibinfo {author} {\bibfnamefont {S.}~\bibnamefont {Poncet}}, \ and\
  \bibinfo {author} {\bibfnamefont {J.-F.}\ \bibnamefont {Pinton}},\ }\bibfield
   {title} {\enquote {\bibinfo {title} {Force measurements on rising
  bubbles},}\ }\href@noop {} {\bibfield  {journal} {\bibinfo  {journal} {J.
  Fluid Mech.}\ }\textbf {\bibinfo {volume} {569}},\ \bibinfo {pages} {51--60}
  (\bibinfo {year} {2006})}\BibitemShut {NoStop}%
\bibitem [{\citenamefont {Zenit}\ and\ \citenamefont
  {Magnaudet}(2008)}]{zenit2008path}%
  \BibitemOpen
  \bibfield  {author} {\bibinfo {author} {\bibfnamefont {R.}~\bibnamefont
  {Zenit}}\ and\ \bibinfo {author} {\bibfnamefont {J.}~\bibnamefont
  {Magnaudet}},\ }\bibfield  {title} {\enquote {\bibinfo {title} {Path
  instability of rising spheroidal air bubbles: a shape-controlled process},}\
  }\href@noop {} {\bibfield  {journal} {\bibinfo  {journal} {Phys. Fluids}\
  }\textbf {\bibinfo {volume} {20}},\ \bibinfo {pages} {061702} (\bibinfo
  {year} {2008})}\BibitemShut {NoStop}%
\bibitem [{\citenamefont {Veldhuis}, \citenamefont {Biesheuvel},\ and\
  \citenamefont {Van~Wijngaarden}(2008)}]{veldhuis2008shape}%
  \BibitemOpen
  \bibfield  {author} {\bibinfo {author} {\bibfnamefont {C.}~\bibnamefont
  {Veldhuis}}, \bibinfo {author} {\bibfnamefont {A.}~\bibnamefont
  {Biesheuvel}}, \ and\ \bibinfo {author} {\bibfnamefont {L.}~\bibnamefont
  {Van~Wijngaarden}},\ }\bibfield  {title} {\enquote {\bibinfo {title} {Shape
  oscillations on bubbles rising in clean and in tap water},}\ }\href@noop {}
  {\bibfield  {journal} {\bibinfo  {journal} {Phys. Fluids}\ }\textbf {\bibinfo
  {volume} {20}},\ \bibinfo {pages} {040705} (\bibinfo {year}
  {2008})}\BibitemShut {NoStop}%
\bibitem [{\citenamefont {Legendre}, \citenamefont {Zenit},\ and\ \citenamefont
  {Velez-Cordero}(2012)}]{Legendre2012}%
  \BibitemOpen
  \bibfield  {author} {\bibinfo {author} {\bibfnamefont {D.}~\bibnamefont
  {Legendre}}, \bibinfo {author} {\bibfnamefont {R.}~\bibnamefont {Zenit}}, \
  and\ \bibinfo {author} {\bibfnamefont {J.~R.}\ \bibnamefont
  {Velez-Cordero}},\ }\bibfield  {title} {\enquote {\bibinfo {title} {On the
  deformation of gas bubbles in liquids},}\ }\href@noop {} {\bibfield
  {journal} {\bibinfo  {journal} {Phys. Fluids}\ }\textbf {\bibinfo {volume}
  {24}},\ \bibinfo {pages} {043303} (\bibinfo {year} {2012})}\BibitemShut
  {NoStop}%
\bibitem [{\citenamefont {Aoyama}\ \emph {et~al.}(2016)\citenamefont {Aoyama},
  \citenamefont {Hayashi}, \citenamefont {Hosokawa},\ and\ \citenamefont
  {Tomiyama}}]{Aoyama2016}%
  \BibitemOpen
  \bibfield  {author} {\bibinfo {author} {\bibfnamefont {S.}~\bibnamefont
  {Aoyama}}, \bibinfo {author} {\bibfnamefont {K.}~\bibnamefont {Hayashi}},
  \bibinfo {author} {\bibfnamefont {S.}~\bibnamefont {Hosokawa}}, \ and\
  \bibinfo {author} {\bibfnamefont {A.}~\bibnamefont {Tomiyama}},\ }\bibfield
  {title} {\enquote {\bibinfo {title} {Shapes of ellipsoidal bubbles in
  infinite stagnant liquids},}\ }\href@noop {} {\bibfield  {journal} {\bibinfo
  {journal} {Int. J. of Multiph. Flow}\ }\textbf {\bibinfo {volume} {79}},\
  \bibinfo {pages} {23--30} (\bibinfo {year} {2016})}\BibitemShut {NoStop}%
\bibitem [{\citenamefont {Rastello}, \citenamefont {Marié},\ and\
  \citenamefont {Lance}(2017)}]{Rastello2017}%
  \BibitemOpen
  \bibfield  {author} {\bibinfo {author} {\bibfnamefont {M.}~\bibnamefont
  {Rastello}}, \bibinfo {author} {\bibfnamefont {J.~L.}\ \bibnamefont
  {Marié}}, \ and\ \bibinfo {author} {\bibfnamefont {M.}~\bibnamefont
  {Lance}},\ }\bibfield  {title} {\enquote {\bibinfo {title} {Clean versus
  contaminated bubbles in a solid-body rotating flow},}\ }\href@noop {}
  {\bibfield  {journal} {\bibinfo  {journal} {J. Fluid Mech.}\ }\textbf
  {\bibinfo {volume} {831}},\ \bibinfo {pages} {592--617} (\bibinfo {year}
  {2017})}\BibitemShut {NoStop}%
\bibitem [{\citenamefont {Agrawal}\ \emph {et~al.}(2021)\citenamefont
  {Agrawal}, \citenamefont {Gaurav}, \citenamefont {Karri},\ and\ \citenamefont
  {Sahu}}]{agrawal2021experimental}%
  \BibitemOpen
  \bibfield  {author} {\bibinfo {author} {\bibfnamefont {M.}~\bibnamefont
  {Agrawal}}, \bibinfo {author} {\bibfnamefont {A.}~\bibnamefont {Gaurav}},
  \bibinfo {author} {\bibfnamefont {B.}~\bibnamefont {Karri}}, \ and\ \bibinfo
  {author} {\bibfnamefont {K.~C.}\ \bibnamefont {Sahu}},\ }\bibfield  {title}
  {\enquote {\bibinfo {title} {An experimental study of two identical air
  bubbles rising side-by-side in water},}\ }\href@noop {} {\bibfield  {journal}
  {\bibinfo  {journal} {Phys.Fluids}\ }\textbf {\bibinfo {volume} {33}}
  (\bibinfo {year} {2021})}\BibitemShut {NoStop}%
\bibitem [{\citenamefont {Kure}\ \emph {et~al.}(2021)\citenamefont {Kure},
  \citenamefont {Jakobsen}, \citenamefont {La~Forgia},\ and\ \citenamefont
  {Solsvik}}]{kure2021experimental}%
  \BibitemOpen
  \bibfield  {author} {\bibinfo {author} {\bibfnamefont {I.~K.}\ \bibnamefont
  {Kure}}, \bibinfo {author} {\bibfnamefont {H.~A.}\ \bibnamefont {Jakobsen}},
  \bibinfo {author} {\bibfnamefont {N.}~\bibnamefont {La~Forgia}}, \ and\
  \bibinfo {author} {\bibfnamefont {J.}~\bibnamefont {Solsvik}},\ }\bibfield
  {title} {\enquote {\bibinfo {title} {Experimental investigation of single
  bubbles rising in stagnant liquid: Statistical analysis and image
  processing},}\ }\href@noop {} {\bibfield  {journal} {\bibinfo  {journal}
  {Phys.Fluids}\ }\textbf {\bibinfo {volume} {33}} (\bibinfo {year}
  {2021})}\BibitemShut {NoStop}%
\bibitem [{\citenamefont {She}\ \emph {et~al.}(2021)\citenamefont {She},
  \citenamefont {Gao}, \citenamefont {Zuo}, \citenamefont {Liao}, \citenamefont
  {Zhao}, \citenamefont {Zhang}, \citenamefont {Nie},\ and\ \citenamefont
  {Shao}}]{She2021}%
  \BibitemOpen
  \bibfield  {author} {\bibinfo {author} {\bibfnamefont {W.~X.}\ \bibnamefont
  {She}}, \bibinfo {author} {\bibfnamefont {Q.}~\bibnamefont {Gao}}, \bibinfo
  {author} {\bibfnamefont {Z.~Y.}\ \bibnamefont {Zuo}}, \bibinfo {author}
  {\bibfnamefont {X.~W.}\ \bibnamefont {Liao}}, \bibinfo {author}
  {\bibfnamefont {L.}~\bibnamefont {Zhao}}, \bibinfo {author} {\bibfnamefont
  {L.~X.}\ \bibnamefont {Zhang}}, \bibinfo {author} {\bibfnamefont {D.~M.}\
  \bibnamefont {Nie}}, \ and\ \bibinfo {author} {\bibfnamefont {X.~M.}\
  \bibnamefont {Shao}},\ }\bibfield  {title} {\enquote {\bibinfo {title}
  {Experimental study on a zigzagging bubble using tomographic particle image
  velocimetry with shadow image reconstruction},}\ }\href {\doibase
  10.13140/RG.2.2.14881.33125} {\bibfield  {journal} {\bibinfo  {journal}
  {Phys.Fluids}\ }\textbf {\bibinfo {volume} {33}},\ \bibinfo {pages}
  {083313} (\bibinfo {year} {2021})}\BibitemShut {NoStop}%
\bibitem [{\citenamefont {Liu}\ \emph {et~al.}(2022)\citenamefont {Liu},
  \citenamefont {Cong}, \citenamefont {Song}, \citenamefont {Wu},\ and\
  \citenamefont {Chen}}]{liu2022experimental}%
  \BibitemOpen
  \bibfield  {author} {\bibinfo {author} {\bibfnamefont {J.}~\bibnamefont
  {Liu}}, \bibinfo {author} {\bibfnamefont {S.}~\bibnamefont {Cong}}, \bibinfo
  {author} {\bibfnamefont {Y.}~\bibnamefont {Song}}, \bibinfo {author}
  {\bibfnamefont {D.}~\bibnamefont {Wu}}, \ and\ \bibinfo {author}
  {\bibfnamefont {S.}~\bibnamefont {Chen}},\ }\bibfield  {title} {\enquote
  {\bibinfo {title} {Experimental study on asymmetric bubbles rising in water:
  Morphology and acoustic signature},}\ }\href@noop {} {\bibfield  {journal}
  {\bibinfo  {journal} {Phys.Fluids}\ }\textbf {\bibinfo {volume} {34}}
  (\bibinfo {year} {2022})}\BibitemShut {NoStop}%
\bibitem [{\citenamefont {Mougin}\ and\ \citenamefont
  {Magnaudet}(2002)}]{Mougin2002}%
  \BibitemOpen
  \bibfield  {author} {\bibinfo {author} {\bibfnamefont {G.}~\bibnamefont
  {Mougin}}\ and\ \bibinfo {author} {\bibfnamefont {J.}~\bibnamefont
  {Magnaudet}},\ }\bibfield  {title} {\enquote {\bibinfo {title} {The
  generalized kirchhoff equations and their application to the interaction
  between a rigid body and an arbitrary time-dependent viscous flow},}\
  }\href@noop {} {\bibfield  {journal} {\bibinfo  {journal} {Int. J. of
  Multiph. Flow}\ }\textbf {\bibinfo {volume} {28}},\ \bibinfo {pages}
  {1837--1851} (\bibinfo {year} {2002})}\BibitemShut {NoStop}%
\bibitem [{\citenamefont {Magnaudet}(2003)}]{Magnaudet2003}%
  \BibitemOpen
  \bibfield  {author} {\bibinfo {author} {\bibfnamefont {J.}~\bibnamefont
  {Magnaudet}},\ }\bibfield  {title} {\enquote {\bibinfo {title} {Small
  inertial effects on a spherical bubble, drop or particle moving near a wall
  in a time-dependent linear flow},}\ }\href@noop {} {\bibfield  {journal}
  {\bibinfo  {journal} {J. Fluid Mech.}\ }\textbf {\bibinfo {volume} {485}},\
  \bibinfo {pages} {115--142} (\bibinfo {year} {2003})}\BibitemShut {NoStop}%
\bibitem [{\citenamefont {Sugiyama}\ and\ \citenamefont
  {Takemura}(2010)}]{sugiyama2010lateral}%
  \BibitemOpen
  \bibfield  {author} {\bibinfo {author} {\bibfnamefont {K.}~\bibnamefont
  {Sugiyama}}\ and\ \bibinfo {author} {\bibfnamefont {F.}~\bibnamefont
  {Takemura}},\ }\bibfield  {title} {\enquote {\bibinfo {title} {On the lateral
  migration of a slightly deformed bubble rising near a vertical plane wall},}\
  }\href@noop {} {\bibfield  {journal} {\bibinfo  {journal} {J. Fluid Mech.}\
  }\textbf {\bibinfo {volume} {662}},\ \bibinfo {pages} {209--231} (\bibinfo
  {year} {2010})}\BibitemShut {NoStop}%
\bibitem [{\citenamefont {Sugioka}\ and\ \citenamefont
  {Tsukada}(2015)}]{sugioka2015direct}%
  \BibitemOpen
  \bibfield  {author} {\bibinfo {author} {\bibfnamefont {K.}~\bibnamefont
  {Sugioka}}\ and\ \bibinfo {author} {\bibfnamefont {T.}~\bibnamefont
  {Tsukada}},\ }\bibfield  {title} {\enquote {\bibinfo {title} {Direct
  numerical simulations of drag and lift forces acting on a spherical bubble
  near a plane wall},}\ }\href@noop {} {\bibfield  {journal} {\bibinfo
  {journal} {Int. J. Multiph. Flow}\ }\textbf {\bibinfo {volume} {71}},\
  \bibinfo {pages} {32--37} (\bibinfo {year} {2015})}\BibitemShut {NoStop}%
\bibitem [{\citenamefont {Zhang}\ \emph {et~al.}(2020)\citenamefont {Zhang},
  \citenamefont {Dabiri}, \citenamefont {Chen},\ and\ \citenamefont
  {You}}]{Zhang2020}%
  \BibitemOpen
  \bibfield  {author} {\bibinfo {author} {\bibfnamefont {Y.}~\bibnamefont
  {Zhang}}, \bibinfo {author} {\bibfnamefont {S.}~\bibnamefont {Dabiri}},
  \bibinfo {author} {\bibfnamefont {K.}~\bibnamefont {Chen}}, \ and\ \bibinfo
  {author} {\bibfnamefont {Y.}~\bibnamefont {You}},\ }\bibfield  {title}
  {\enquote {\bibinfo {title} {An initially spherical bubble rising near a
  vertical wall},}\ }\href@noop {} {\bibfield  {journal} {\bibinfo  {journal}
  {Int. J. Heat Fluid Flow}\ }\textbf {\bibinfo {volume} {85}},\ \bibinfo
  {pages} {108649} (\bibinfo {year} {2020})}\BibitemShut {NoStop}%
\bibitem [{\citenamefont {Hasan}\ and\ \citenamefont
  {Hasan}(2022)}]{Hasan2022}%
  \BibitemOpen
  \bibfield  {author} {\bibinfo {author} {\bibfnamefont {S.~M.~M.}\
  \bibnamefont {Hasan}}\ and\ \bibinfo {author} {\bibfnamefont {A.~B. M.~T.}\
  \bibnamefont {Hasan}},\ }\bibfield  {title} {\enquote {\bibinfo {title}
  {{Migration dynamics of an initially spherical deformable bubble in the
  vicinity of a corner}},}\ }\href {\doibase 10.1063/5.0115162} {\bibfield
  {journal} {\bibinfo  {journal} {Phys.Fluids}\ }\textbf {\bibinfo
  {volume} {34}},\ \bibinfo {pages} {112119} (\bibinfo {year}
  {2022})}\BibitemShut {NoStop}%
\bibitem [{\citenamefont {Yan}\ \emph {et~al.}(2022)\citenamefont {Yan},
  \citenamefont {Zhang}, \citenamefont {Liao}, \citenamefont {Zhang},
  \citenamefont {Zhou},\ and\ \citenamefont {Liu}}]{Yan2022}%
  \BibitemOpen
  \bibfield  {author} {\bibinfo {author} {\bibfnamefont {H.}~\bibnamefont
  {Yan}}, \bibinfo {author} {\bibfnamefont {H.}~\bibnamefont {Zhang}}, \bibinfo
  {author} {\bibfnamefont {Y.}~\bibnamefont {Liao}}, \bibinfo {author}
  {\bibfnamefont {H.}~\bibnamefont {Zhang}}, \bibinfo {author} {\bibfnamefont
  {P.}~\bibnamefont {Zhou}}, \ and\ \bibinfo {author} {\bibfnamefont
  {L.}~\bibnamefont {Liu}},\ }\bibfield  {title} {\enquote {\bibinfo {title} {A
  single bubble rising in the vicinity of a vertical wall: A numerical study
  based on volume of fluid method},}\ }\href@noop {} {\bibfield  {journal}
  {\bibinfo  {journal} {Ocean Eng.}\ }\textbf {\bibinfo {volume} {263}},\
  \bibinfo {pages} {112379} (\bibinfo {year} {2022})}\BibitemShut {NoStop}%
\bibitem [{\citenamefont {Yan}\ \emph {et~al.}(2023)\citenamefont {Yan},
  \citenamefont {Zhang}, \citenamefont {Zhang}, \citenamefont {Liao},\ and\
  \citenamefont {Liu}}]{yan2023three}%
  \BibitemOpen
  \bibfield  {author} {\bibinfo {author} {\bibfnamefont {H.}~\bibnamefont
  {Yan}}, \bibinfo {author} {\bibfnamefont {H.-Y.}\ \bibnamefont {Zhang}},
  \bibinfo {author} {\bibfnamefont {H.-M.}\ \bibnamefont {Zhang}}, \bibinfo
  {author} {\bibfnamefont {Y.-X.}\ \bibnamefont {Liao}}, \ and\ \bibinfo
  {author} {\bibfnamefont {L.}~\bibnamefont {Liu}},\ }\bibfield  {title}
  {\enquote {\bibinfo {title} {Three-dimensional dynamics of a single bubble
  rising near a vertical wall: Paths and wakes},}\ }\href@noop {} {\bibfield
  {journal} {\bibinfo  {journal} {Pet. Sci.}\ }\textbf {\bibinfo {volume} {In
  Press}} (\bibinfo {year} {2023})}\BibitemShut {NoStop}%
\bibitem [{\citenamefont {Takemura}\ \emph {et~al.}(2002)\citenamefont
  {Takemura}, \citenamefont {Takagi}, \citenamefont {Magnaudet},\ and\
  \citenamefont {Matsumoto}}]{Takemura2002}%
  \BibitemOpen
  \bibfield  {author} {\bibinfo {author} {\bibfnamefont {F.}~\bibnamefont
  {Takemura}}, \bibinfo {author} {\bibfnamefont {S.}~\bibnamefont {Takagi}},
  \bibinfo {author} {\bibfnamefont {J.}~\bibnamefont {Magnaudet}}, \ and\
  \bibinfo {author} {\bibfnamefont {Y.}~\bibnamefont {Matsumoto}},\ }\bibfield
  {title} {\enquote {\bibinfo {title} {Drag and lift forces on a bubble rising
  near a vertical wall in a viscous liquid},}\ }\href@noop {} {\bibfield
  {journal} {\bibinfo  {journal} {J. Fluid Mech.}\ }\textbf {\bibinfo {volume}
  {461}},\ \bibinfo {pages} {277--300} (\bibinfo {year} {2002})}\BibitemShut
  {NoStop}%
\bibitem [{\citenamefont {Takemura}\ and\ \citenamefont
  {Magnaudet}(2003)}]{Takemura2003}%
  \BibitemOpen
  \bibfield  {author} {\bibinfo {author} {\bibfnamefont {F.}~\bibnamefont
  {Takemura}}\ and\ \bibinfo {author} {\bibfnamefont {J.}~\bibnamefont
  {Magnaudet}},\ }\bibfield  {title} {\enquote {\bibinfo {title} {The
  transverse force on clean and contaminated bubbles rising near a vertical
  wall at moderate {R}eynolds number},}\ }\href@noop {} {\bibfield  {journal}
  {\bibinfo  {journal} {J. Fluid Mech.}\ }\textbf {\bibinfo {volume} {495}},\
  \bibinfo {pages} {235--253} (\bibinfo {year} {2003})}\BibitemShut {NoStop}%
\bibitem [{\citenamefont {Jeong}\ and\ \citenamefont {Park}(2015)}]{Jeong2015}%
  \BibitemOpen
  \bibfield  {author} {\bibinfo {author} {\bibfnamefont {H.}~\bibnamefont
  {Jeong}}\ and\ \bibinfo {author} {\bibfnamefont {H.}~\bibnamefont {Park}},\
  }\bibfield  {title} {\enquote {\bibinfo {title} {Near-wall rising behaviour
  of a deformable bubble at high reynolds number},}\ }\href@noop {} {\bibfield
  {journal} {\bibinfo  {journal} {J. Fluid Mech.}\ }\textbf {\bibinfo {volume}
  {771}},\ \bibinfo {pages} {564--594} (\bibinfo {year} {2015})}\BibitemShut
  {NoStop}%
\bibitem [{\citenamefont {Lee}\ and\ \citenamefont {Park}(2017)}]{Lee2017}%
  \BibitemOpen
  \bibfield  {author} {\bibinfo {author} {\bibfnamefont {J.}~\bibnamefont
  {Lee}}\ and\ \bibinfo {author} {\bibfnamefont {H.}~\bibnamefont {Park}},\
  }\bibfield  {title} {\enquote {\bibinfo {title} {Wake structures behind an
  oscillating bubble rising close to a vertical wall},}\ }\href@noop {}
  {\bibfield  {journal} {\bibinfo  {journal} {Int. J. Multiph. Flow}\ }\textbf
  {\bibinfo {volume} {91}},\ \bibinfo {pages} {225--242} (\bibinfo {year}
  {2017})}\BibitemShut {NoStop}%
\bibitem [{\citenamefont {Cano-Lozano}, \citenamefont {Bohorquez},\ and\
  \citenamefont {Martínez-Bazán}(2013)}]{Cano-Lozano2013}%
  \BibitemOpen
  \bibfield  {author} {\bibinfo {author} {\bibfnamefont {J.~C.}\ \bibnamefont
  {Cano-Lozano}}, \bibinfo {author} {\bibfnamefont {P.}~\bibnamefont
  {Bohorquez}}, \ and\ \bibinfo {author} {\bibfnamefont {C.}~\bibnamefont
  {Martínez-Bazán}},\ }\bibfield  {title} {\enquote {\bibinfo {title} {Wake
  instability of a fixed axisymmetric bubble of realistic shape},}\ }\href@noop
  {} {\bibfield  {journal} {\bibinfo  {journal} {Int. J. of Multiph. Flow}\
  }\textbf {\bibinfo {volume} {51}},\ \bibinfo {pages} {11--21} (\bibinfo
  {year} {2013})}\BibitemShut {NoStop}%
\bibitem [{\citenamefont {Cano-Lozano}\ \emph
  {et~al.}(2016{\natexlab{b}})\citenamefont {Cano-Lozano}, \citenamefont
  {Tchoufag}, \citenamefont {Magnaudet},\ and\ \citenamefont
  {Mart{\'\i}nez-Baz{\'a}n}}]{Cano-Lozano2016a}%
  \BibitemOpen
  \bibfield  {author} {\bibinfo {author} {\bibfnamefont {J.~C.}\ \bibnamefont
  {Cano-Lozano}}, \bibinfo {author} {\bibfnamefont {J.}~\bibnamefont
  {Tchoufag}}, \bibinfo {author} {\bibfnamefont {J.}~\bibnamefont {Magnaudet}},
  \ and\ \bibinfo {author} {\bibfnamefont {C.}~\bibnamefont
  {Mart{\'\i}nez-Baz{\'a}n}},\ }\bibfield  {title} {\enquote {\bibinfo {title}
  {A global stability approach to wake and path instabilities of nearly oblate
  spheroidal rising bubbles},}\ }\href@noop {} {\bibfield  {journal} {\bibinfo
  {journal} {Phys. Fluids}\ }\textbf {\bibinfo {volume} {28}},\ \bibinfo
  {pages} {014102} (\bibinfo {year} {2016}{\natexlab{b}})}\BibitemShut
  {NoStop}%
\bibitem [{\citenamefont {Rubio-Rubio}\ \emph {et~al.}(2021)\citenamefont
  {Rubio-Rubio}, \citenamefont {Bolaños-Jiménez}, \citenamefont
  {Martínez-Bazán}, \citenamefont {Muñoz-Hervás},\ and\ \citenamefont
  {Sevilla}}]{Rubio-Rubio2021}%
  \BibitemOpen
  \bibfield  {author} {\bibinfo {author} {\bibfnamefont {M.}~\bibnamefont
  {Rubio-Rubio}}, \bibinfo {author} {\bibfnamefont {R.}~\bibnamefont
  {Bolaños-Jiménez}}, \bibinfo {author} {\bibfnamefont {C.}~\bibnamefont
  {Martínez-Bazán}}, \bibinfo {author} {\bibfnamefont {J.~C.}\ \bibnamefont
  {Muñoz-Hervás}}, \ and\ \bibinfo {author} {\bibfnamefont {A.}~\bibnamefont
  {Sevilla}},\ }\bibfield  {title} {\enquote {\bibinfo {title}
  {Superhydrophobic substrates allow the generation of giant quasi-static
  bubbles},}\ }\href@noop {} {\bibfield  {journal} {\bibinfo  {journal} {J.
  Fluid Mech.}\ }\textbf {\bibinfo {volume} {912}},\ \bibinfo {pages} {1--14}
  (\bibinfo {year} {2021})}\BibitemShut {NoStop}%
\bibitem [{\citenamefont {Dorsey}(1940)}]{dorsey_glyc-water}%
  \BibitemOpen
  \bibfield  {author} {\bibinfo {author} {\bibfnamefont {N.~E.}\ \bibnamefont
  {Dorsey}},\ }\href@noop {} {\emph {\bibinfo {title} {Properties of Ordinary
  Water-Substance in All Its Phases: Water-Vapor}}}\ (\bibinfo  {publisher}
  {Reinhold Publishing Company},\ \bibinfo {year} {1940})\BibitemShut {NoStop}%
\bibitem [{\citenamefont {{Glycerine Producers’
  Association}}(1963)}]{glycerine}%
  \BibitemOpen
  \bibfield  {author} {\bibinfo {author} {\bibnamefont {{Glycerine Producers’
  Association}}},\ }\href@noop {} {\enquote {\bibinfo {title} {Physical
  properties of glycerine and its solutions},}\ }\bibinfo {type} {Tech. Rep.}\
  (\bibinfo  {institution} {Glycerine Producers’ Association, New York},\
  \bibinfo {year} {1963})\BibitemShut {NoStop}%
\bibitem [{\citenamefont {Streit}\ \emph {et~al.}(2005)\citenamefont {Streit},
  \citenamefont {Sprung}, \citenamefont {Gutt},\ and\ \citenamefont
  {Tolan}}]{Streit_glyc}%
  \BibitemOpen
  \bibfield  {author} {\bibinfo {author} {\bibfnamefont {S.}~\bibnamefont
  {Streit}}, \bibinfo {author} {\bibfnamefont {M.}~\bibnamefont {Sprung}},
  \bibinfo {author} {\bibfnamefont {C.}~\bibnamefont {Gutt}}, \ and\ \bibinfo
  {author} {\bibfnamefont {M.}~\bibnamefont {Tolan}},\ }\bibfield  {title}
  {\enquote {\bibinfo {title} {The surface roughness of water/glycerol mixtures
  investigated by x-ray reflectivity},}\ }\href@noop {} {\bibfield  {journal}
  {\bibinfo  {journal} {Physica B Condens. Matter}\ }\textbf {\bibinfo {volume}
  {357}},\ \bibinfo {pages} {110--114} (\bibinfo {year} {2005})}\BibitemShut
  {NoStop}%
\bibitem [{\citenamefont {Spann}\ \emph {et~al.}(2021)\citenamefont {Spann},
  \citenamefont {Hancock}, \citenamefont {Rostami}, \citenamefont {Platt},
  \citenamefont {Rusyniak}, \citenamefont {Sundar}, \citenamefont {Lau},\ and\
  \citenamefont {Pithawalla}}]{Spann_glyc}%
  \BibitemOpen
  \bibfield  {author} {\bibinfo {author} {\bibfnamefont {A.~P.}\ \bibnamefont
  {Spann}}, \bibinfo {author} {\bibfnamefont {M.~J.}\ \bibnamefont {Hancock}},
  \bibinfo {author} {\bibfnamefont {A.~A.}\ \bibnamefont {Rostami}}, \bibinfo
  {author} {\bibfnamefont {S.~P.}\ \bibnamefont {Platt}}, \bibinfo {author}
  {\bibfnamefont {M.~J.}\ \bibnamefont {Rusyniak}}, \bibinfo {author}
  {\bibfnamefont {R.~S.}\ \bibnamefont {Sundar}}, \bibinfo {author}
  {\bibfnamefont {R.~W.}\ \bibnamefont {Lau}}, \ and\ \bibinfo {author}
  {\bibfnamefont {Y.~B.}\ \bibnamefont {Pithawalla}},\ }\bibfield  {title}
  {\enquote {\bibinfo {title} {Viscosity model for liquid mixtures of propylene
  glycol, glycerol, and water},}\ }\href@noop {} {\bibfield  {journal}
  {\bibinfo  {journal} {Ind. Eng. Chem. Res.}\ }\textbf {\bibinfo {volume}
  {60}},\ \bibinfo {pages} {670--677} (\bibinfo {year} {2021})}\BibitemShut
  {NoStop}%
\bibitem [{\citenamefont {Otsu}(1979)}]{Otsu1979}%
  \BibitemOpen
  \bibfield  {author} {\bibinfo {author} {\bibfnamefont {N.}~\bibnamefont
  {Otsu}},\ }\bibfield  {title} {\enquote {\bibinfo {title} {A threshold
  selection method from gray-level histograms},}\ }\href@noop {} {\bibfield
  {journal} {\bibinfo  {journal} {IEEE Trans. Syst. Man Cybern.}\ }\textbf
  {\bibinfo {volume} {9}},\ \bibinfo {pages} {62--66} (\bibinfo {year}
  {1979})}\BibitemShut {NoStop}%
\bibitem [{\citenamefont {Bradley}\ and\ \citenamefont
  {Roth}(2007)}]{Bradley2007}%
  \BibitemOpen
  \bibfield  {author} {\bibinfo {author} {\bibfnamefont {D.}~\bibnamefont
  {Bradley}}\ and\ \bibinfo {author} {\bibfnamefont {G.}~\bibnamefont {Roth}},\
  }\bibfield  {title} {\enquote {\bibinfo {title} {Adaptive thresholding using
  the integral image},}\ }\href@noop {} {\bibfield  {journal} {\bibinfo
  {journal} {J. Graphics Tools}\ }\textbf {\bibinfo {volume} {12}},\ \bibinfo
  {pages} {13--21} (\bibinfo {year} {2007})}\BibitemShut {NoStop}%
\bibitem [{\citenamefont {Ho}(1983)}]{Ho1983}%
  \BibitemOpen
  \bibfield  {author} {\bibinfo {author} {\bibfnamefont {C.-S.}\ \bibnamefont
  {Ho}},\ }\bibfield  {title} {\enquote {\bibinfo {title} {Precision of digital
  vision systems},}\ }\href@noop {} {\bibfield  {journal} {\bibinfo  {journal}
  {EEE Trans. Pattern Anal. Mach. Intell.}\ }\textbf {\bibinfo {volume}
  {PAMI-5}},\ \bibinfo {pages} {593--601} (\bibinfo {year} {1983})}\BibitemShut
  {NoStop}%
\bibitem [{\citenamefont {Leal}(1980)}]{leal1980particle}%
  \BibitemOpen
  \bibfield  {author} {\bibinfo {author} {\bibfnamefont {L.}~\bibnamefont
  {Leal}},\ }\bibfield  {title} {\enquote {\bibinfo {title} {Particle motions
  in a viscous fluid},}\ }\href@noop {} {\bibfield  {journal} {\bibinfo
  {journal} {Annu. Rev. Fluid Mech.}\ }\textbf {\bibinfo {volume} {12}},\
  \bibinfo {pages} {435--476} (\bibinfo {year} {1980})}\BibitemShut {NoStop}%
\bibitem [{\citenamefont {Rastello}, \citenamefont {Marié},\ and\
  \citenamefont {Lance}(2011)}]{Rastello2011}%
  \BibitemOpen
  \bibfield  {author} {\bibinfo {author} {\bibfnamefont {M.}~\bibnamefont
  {Rastello}}, \bibinfo {author} {\bibfnamefont {J.-L.}\ \bibnamefont
  {Marié}}, \ and\ \bibinfo {author} {\bibfnamefont {M.}~\bibnamefont
  {Lance}},\ }\bibfield  {title} {\enquote {\bibinfo {title} {Drag and lift
  forces on clean spherical and ellipsoidal bubbles in a solid-body rotating
  flow},}\ }\href@noop {} {\bibfield  {journal} {\bibinfo  {journal} {J. Fluid
  Mech.}\ }\textbf {\bibinfo {volume} {682}},\ \bibinfo {pages} {434–459}
  (\bibinfo {year} {2011})}\BibitemShut {NoStop}%
\bibitem [{\citenamefont {Br{\"u}cker}(1999)}]{brucker1999structure}%
  \BibitemOpen
  \bibfield  {author} {\bibinfo {author} {\bibfnamefont {C.}~\bibnamefont
  {Br{\"u}cker}},\ }\bibfield  {title} {\enquote {\bibinfo {title} {Structure
  and dynamics of the wake of bubbles and its relevance for bubble
  interaction},}\ }\href@noop {} {\bibfield  {journal} {\bibinfo  {journal}
  {Phys. Fluids}\ }\textbf {\bibinfo {volume} {11}},\ \bibinfo {pages}
  {1781--1796} (\bibinfo {year} {1999})}\BibitemShut {NoStop}%
\bibitem [{\citenamefont {Zenit}\ and\ \citenamefont
  {Magnaudet}(2009)}]{zenit2009measurements}%
  \BibitemOpen
  \bibfield  {author} {\bibinfo {author} {\bibfnamefont {R.}~\bibnamefont
  {Zenit}}\ and\ \bibinfo {author} {\bibfnamefont {J.}~\bibnamefont
  {Magnaudet}},\ }\bibfield  {title} {\enquote {\bibinfo {title} {Measurements
  of the streamwise vorticity in the wake of an oscillating bubble},}\
  }\href@noop {} {\bibfield  {journal} {\bibinfo  {journal} {Int. J. Multiph.
  Flow}\ }\textbf {\bibinfo {volume} {35}},\ \bibinfo {pages} {195--203}
  (\bibinfo {year} {2009})}\BibitemShut {NoStop}%
\end{thebibliography}
\end{document}